\newlength{\extralineskip}

\documentstyle[draft]{article}
\makeatletter
\newdimen\normalarrayskip              
\newdimen\minarrayskip                 
\normalarrayskip\baselineskip
\minarrayskip\jot
\newif\ifold             \oldtrue            \def\new{\oldfalse}
\def\arraymode{\ifold\relax\else\displaystyle\fi} 
\def\eqnumphantom{\phantom{(\theequation)}}     
\def\@arrayskip{\ifold\baselineskip\z@\lineskip\z@
     \else
     \baselineskip\minarrayskip\lineskip2\minarrayskip\fi}
\def\@arrayclassz{\ifcase \@lastchclass \@acolampacol \or
\@ampacol \or \or \or \@addamp \or
   \@acolampacol \or \@firstampfalse \@acol \fi
\edef\@preamble{\@preamble
  \ifcase \@chnum
     \hfil$\relax\arraymode\@sharp$\hfil
     \or $\relax\arraymode\@sharp$\hfil
     \or \hfil$\relax\arraymode\@sharp$\fi}}
\def\@array[#1]#2{\setbox\@arstrutbox=\hbox{\vrule
     height\arraystretch \ht\strutbox
     depth\arraystretch \dp\strutbox
     width\z@}\@mkpream{#2}\edef\@preamble{\halign \noexpand\@halignto
\bgroup \tabskip\z@ \@arstrut \@preamble \tabskip\z@ \cr}%
\let\@startpbox\@@startpbox \let\@endpbox\@@endpbox
  \if #1t\vtop \else \if#1b\vbox \else \vcenter \fi\fi
  \bgroup \let\par\relax
  \let\@sharp##\let\protect\relax
  \@arrayskip\@preamble}
\def\eqnarray{\stepcounter{equation}%
              \let\@currentlabel=\theequation
              \global\@eqnswtrue
              \global\@eqcnt\z@
              \tabskip\@centering
              \let\\=\@eqncr
              $$%
 \halign to \displaywidth\bgroup
    \eqnumphantom\@eqnsel\hskip\@centering
    $\displaystyle \tabskip\z@ {##}$%
    &\global\@eqcnt\@ne \hskip 2\arraycolsep
         $\displaystyle\arraymode{##}$\hfil
    &\global\@eqcnt\tw@ \hskip 2\arraycolsep
         $\displaystyle\tabskip\z@{##}$\hfil
         \tabskip\@centering
    &{##}\tabskip\z@\cr}
\makeatother
\def\tr#1{{\rm tr}\kern-3pt\left[#1\right]}

\def\nn{\nonumber}

\def\beq{\begin{equation}}
\def\eeq{\end{equation}}
\def\be{\beq\new\begin{array}{c}}
\def\ee{\end{array}\eeq}

\newcommand{\sect}[1]{\setcounter{equation}{0}\section{#1}}
\renewcommand{\theequation}{\thesection.\arabic{equation}}

\begin{document}

\begin{titlepage}
\setcounter{footnote}0
\begin{center}

\phantom . \hfill ITEP-M6/95 \\
\phantom . \hfill OU-HET-230 \\
\vspace{0.5in}

{\Large{PREPOTENTIAL and the SEIBERG-WITTEN THEORY}}
\\[.2in]

{\it H.Itoyama\footnote{Departement of Physics, Graduate School of Science,
Osaka University, Toyonaka, Osaka 560, Japan.
E-mail address: itoyama@funpth.phys.sci.osaka-u.ac.jp} and
A.Morozov\footnote{117259, ITEP, Moscow,  Russia.
E-mail address: morozov@vxitep.itep.ru}} \\

\end{center}
\bigskip
\bigskip

\centerline{\bf ABSTRACT}
\begin{quotation}
Some basic facts about the prepotential in the SW/Whitham theory
are presented. Consideration begins from the abstract theory
of quasiclassical $\tau$-functions , which uses as
input a family of complex spectral curves with a meromorphic
differential $dS$, subject to the constraint $\partial dS/\partial(moduli)
= \ holomorphic$,  and gives as an output a homogeneous
prepotential on extended moduli space. Then reversed construction
is discussed, which is straightforwardly generalizable from
spectral {\it curves} to certain
complex manifolds of dimension $d >1$ (like $K3$ and $CY$ families).
Finally, examples of particular $N=2$ SUSY gauge models are
considered from the point of view of this formalism.
At the end we discuss similarity between the
$WP^{12}_{1,1,2,2,6}$ -\-Calabi-\-Yau model with $h_{21}=2$
and the $1d$ $SL(2)$ Calogero/Ruijsenaars model, but stop short
of the claim that they belong to the same Whitham
universality class beyond the conifold limit.

\end{quotation}


\end{titlepage}

\tableofcontents  \newpage

\setcounter{footnote}{0}

According to \cite{SW,SW2}
the low-energy effective actions for $4d$ $N=2$ SUSY YM theories
are described in terms of Riemann surfaces (complex curves),
${\cal C}(h_k)$ and certain integrals $S_C = \oint_C dS$ along
non-contractable contours on ${\cal C}$.
The simplest way to summarize these results is in terms of
$1d$ integrable systems \cite{Go}. For a brief review of this approach
and references see \cite{IM}.
The purpose of this paper is to present some details,
concerning the adequate definition and properties of
the differential $dS$.  Most of them are well known to
the experts, see especially \cite{Kr2,D,Go,NT,EY,STY};
our presentation is necessarily close to these references.

\section{Introduction}

There are various starting points which can serve as a motivation
for the study of quasiclassical $\tau$-functions.

The most general one is as follows. Given a classical dynamical system
one can think of  two ways to proceed after exact angle-action variables
are somehow found for it. First, one can quantize the system. Second, one
can average over fast fluctuations of angle-variables and get some
effective slow dynamics on the space of integrals of motion. Though
seemingly different, these are exactly the same problems, at least in the
first approximation (known as non-linear WKB or Bogolubov-Whitham
approximation). Basically, the reason is that quantum wave functions
appear from  averaging along the classical trajectory  - very much in the
spirit of ergodicity theorems.
In the modern string-theory language  the
classical system in question arises after some first-quantized problem
is exactly solved,
its effective action (generating function of all the correlators in the given
background field)  being
a $\tau$-function of some underlying loop-group symmetry. Except for
particular  stringy examples, it is yet rarely
known in any explicit form. The two above mentioned problems
concern deformation
of  classical into quantum symmetry and renormalization group flow to the
low-energy (topological) field theory.
Effective action arising after averaging over fast fluctuations (at the end
point of RG flow)  is somewhat different from original
$\tau$-function, and often much simpler in some respects. Still, in varience
with original one (which is a generating functional of all the
matrix elements of some group), this "quasiclassical $\tau$-function"
- at the present stage of our knowledge - does not have any nice
group-theoretical interpretation. Moreover, there is no clear understanding
what all the examples of quasiclassical $\tau$-functions, that emerge in
different contexts (from application of Whitham method and from quantization
procedures, from the study of KP/Toda hierarchies and that of topological
theories on Calabi-Yau manifolds) have in common (see \cite{Kr2}
and \cite{D} for attempts of ''universal'' definitions).

Emergence of quasiclassical $\tau$-functions in the role of (exponentiated)
prepotentials in exactly
solvable supersymmetric $4d$ theories - though not a surprise from
this general point of view - unavoidably stimulates new interest to these
underinvestigated objects. Moreover, their new appearance in
\cite{SW,SW2} is remarkably similar to old examples, related to KP/Toda
hierarchies. As there, the crucial ingredient is a family of Riemann
surfaces ${\cal C}(h)$ with meromorphic (1,0)-differentials $dS$,
subject to the constraint that  derivatives of 
are holomorphic. Once such family is
introduced, the very existence of prepotential and its homogeneity
- a subject of certain attention recently - are {\it immediate} corollaries.

It is this last  relation - between the differentials $dS$ and
  the prepotentials -
 that is the topic of  the present paper.  We do not go into more details
of above-described general philosophy \cite{UFN}
around the notion of $\tau$-functions
- neither classical nor quasiclassical.  Instead we investigate formally the
corollaries of the {\it definition} of $dS$, just mentioning briefly relations
to various other problems which we meet on the way.
At least one of  the formulations of the prepotential theory allows
for immediate generalization from spectral {\it curves} to complex manifolds
of arbitrary dimension - and thus identify the prepotentials of the
special geometry as (logarithms of) the quasiclassical $\tau$-functions.
In the last sections we go backwards, and study some particular
examples, related to various $N=2$ SUSY gauge theories, showing
how they suit into the general framework.

\section{Notations}

First of all, it is only cohomology class of $dS$ that matters, therefore in
what follows we use ''$\cong$''
instead of ''$=$''  to designate equivalence modulo total derivative.
Futher, let $d\omega_i$, $\ i = 1,\ldots, p$ denote canonical holomorphic
1-differentials on the complex ''spectral'' curve (Riemann surface)
${\cal C}(h_k)$, where $p$ is the genus of
${\cal C}$ and $h_k$ are its moduli (of complex structure), so that
$\oint_{A_i}d\omega_j = \delta_{ij}$, $\oint_{B_i} d\omega_j = {\cal T}_{ij}$,
and ${\cal T}_{ij}(h_k)$ is period matrix.

 From the point of view of original YM theory (refered to in this context as
the ''target space theory'') the type of the curve ${\cal C}$, in particular
its genus  $p$,  is defined
by the gauge group $G$, while $h_k$ are the vacuum
expectation values $\ \frac{1}{k}<{\rm Tr}\Phi^k>\ $
of a scalar field in the adjoint of $G$, which break
original gauge symmetry down to $U(1)^{r_G}$.
 From the point of view of integrability theory, the type of ${\cal C}$
is related to the choice of integrable system, and $h_k$ are its
Hamiltonians (integrals of motion). See \cite{IM} for details.

\subsection{Definition of $dS$}

The meromorphic 1-differential $dS$ on ${\cal C}$ can be
(partly) {\it defined} by the requirement:\footnote{
This definition has motivations both in the $4d$
and integrability theories. In the former case \cite{SW,SW2} it allows one to
make a positive definite Kahlerian metric on moduli space
from $Im\sum_{i=1}^p \delta\oint_{A_i}dS \wedge \delta\oint_{B_i}dS$.
In the latter case it allows to build from $dS$ a solution to  Whitham
hierarchy.
}
\be
\frac{\partial\  dS}{\partial h_k} \cong  \ holomorphic\ {\rm differential},
\label{dSdefor}
\ee
or, in the given basis  $\{d\omega_i\}$,
\be
\frac{\partial\  dS}{\partial h_k} \cong \sum_{i=1}^p
\sigma_{ki}^{(dS)} d\omega_i   \ \ \ \ \ {\rm for\ all\ } h_k.
\label{dSdef}
\ee
Of course, everything here - $dS$, $\sigma$, $d\omega$ - depend on
the moduli $h_k$; the coefficients $\sigma$ depend also on
particular  solution $dS$. The equation (\ref{dSdef}) is somewhat
restrictive.
Normally an object which is holomorphic in a given complex structure,
becomes non-holomorphic in another one. At best, one can adjust
things so that holomorphic objects acquire poles when differentiated
with respect to the moduli. Eq.(\ref{dSdef}) demands quite the opposite
when applied to a meromorphic $dS$:
the variation of complex structure should {\it eliminate} the poles.
The option that allows meromorphic solutions to (\ref{dSdef}) to exist
is that the newly emerging poles can coincide with the zeroes of $dS$,
and thus are canceled. Another requirement is that  residues of the
poles  of $dS$ itself, if any, should be independent of moduli.

Existence of solutions to (\ref{dSdef}) depend on the family of the curves
${\cal C}$, in particular, on the number $K \equiv \#k$
of moduli $h_k$ involved.
If in (\ref{dSdef}) $\# k = K < p = \# i $, then even holomorphic solutions
$dS$
can exist. If $K = p$ -- the case to be actually discussed below
(and associated with pure gauge $N=2$ SUSY theories in $4d$ and with
Toda-chain systems in $1d$) -- there are no holomorphic solutions, but
those with {\it any} non-empty set of poles do exist. If $K > p$ (as in the
case of $N=2$ theory with matter multiplets in the fundamental of $G$
and also for generic families of Riemann surfaces  and integrable systems),
the number of poles of $dS$ is restricted from below.
In what follows we concentrate on the most illuminating case of
$K = p$.

\subsection{Basis in the space of solutions}

Let us introduce a basis in the linear space of solutions to (\ref{dSdef}).
For this, as is usual in the theory of integrable systems, select two points
$\xi_\pm$ on ${\cal C}$ and fix somehow a complex coordinate $\xi$
in the vicinities of $\xi_\pm$.\footnote{
For notational simplicity we assume that coordinate system is common for
the vicinities of $\xi_+$ and $\xi_-$. Actually, this is not a restriction.
Our presentation is mostly in the context of Toda-lattice hierarchy.
In the case of KP, $t_n = 0$ for all $n\leq 0$, and only the puncture at
$\xi_+$ matters. Another important reduction of Toda-lattice identifies
$t_{-n} = (-)^{n-1}t_n$, what can be also described as substitution of the
curve ${\cal C}$ by $\hat{\cal C}$ of which ${\cal C}$ is a double cover.
In coordinates this can be described as
$\frac{1}{\hat\xi - \xi_0} = \frac{1}{\xi - \xi_+} - \frac{1}{\xi - \xi_-}$
(the most familiar formula arises for $\xi_+ = 0$, $\xi_- \xi_0 = \infty$:
$\hat\xi =   \xi + \xi^{-1}$). Instead of reductions of Toda-lattice one
can instead consider generalizations, when there are many punctures
at points $\xi_\alpha$. Associated time-variables are $T_{\alpha,n}$,
$n>0$, together with certain set of zero-time variables.
Additional punctures appear in considerations of $N=2$ SUSY theories
with additional matter supermultiplets.
}
Then define $d\hat\Omega_n$ as a solution to
(\ref{dSdef}), which satisfies additional constraints:

(i) $d\hat\Omega_{\pm n}$ for $n\geq 1$
has a pole of order $n+1$ at $\xi_\pm$ and no other poles;
$d\hat\Omega_0$ has simple poles at both $\xi_+$ and $\xi_-$.

(ii) $d\hat\Omega_{\pm n}(\xi)  =  \pm\left[ (\xi - \xi_\pm)^{-n-1}
 + o(1)\right] d\xi$.

$d\hat\Omega_0$ has simple poles at both $\pm\xi$ with residues $\pm1$.
When $d\hat\Omega_n$ exist for all $n$ - as is the case of
$K = p$, (ii) can be easily fulfilled by triangular transformation
of any basis which satisfies (i).
In our basis eq.(\ref{dSdef}) acquires the form:
\be
 \frac{\partial\  d\hat\Omega_n}{\partial h_k} \cong \sum_{i=1}^p
\sigma_{ki}^{(n)} d\omega_i
\label{dSdef'}
\ee

\subsection{Another basis and solutions to KP/Toda hierarchy}

Requirements (i) and (ii) are just the same as those in the theory of
KP/Toda-hierarchies \cite{Kr}, where {\it instead} of being a solution to
(\ref{dSdef}), the basis element $d\Omega_n$ is requested to satisfy:

(iii)  $\oint_{A_j} d\Omega_n = 0\ \ \ $ for all $A$-cycles.

Then, the $B$-periods of such $d\Omega_n$,
$\ \kappa^{(n)}_j \equiv \oint_{B_j} d\Omega_n$, are
the frequencies,  which - together with the twists made
from $\oint_{B_j} d\omega_i = {\cal T}_{ij}$,
enter the argument of  the theta-function in the
celebrated formula,
\be
\tau(t | h) = e^{f(\alpha_i, t_n | h)}
\Theta\left(  s_{B_j} \left| {\cal T}_{ij}(h)\right.\right), \nn \\
s_{B_j}(\alpha,t|h) = \oint_{B_j}ds =
\oint_{B_j}\left(\sum_{i=1}^p \alpha_id\omega_i + \sum_{n=-\infty}^\infty
t_nd\Omega_n\right) =
\sum_{i=1}^p \alpha_i{\cal T}_{ij} +
\sum_{n=-\infty}^\infty \kappa_j^{(n)}(h) t_n,
\label{KPtau}
\ee
for KP/Toda $\tau$-function, associated with the Riemann surface
${\cal C}$ (a $p$-zone solution).
The main theorem of the KP/Toda theory \cite{Kr} says that
$h_k$ in eq.(\ref{KPtau}) are independent of $\alpha_i, t_n$
(moduli are invariants of KP/Toda flows). Thus $s_{B_j}$
are linear functions of $\alpha_i$ and $t_n$:
$$
\ {\cal T}_{ij} = \frac{\partial s_{B_j}}{\partial \alpha_i}, \ \ \
\ \kappa_j^{(n)} = \frac{\partial s_{B_j}}{\partial t_n},\ \ \
  {\rm and} \ \ \
\ \frac{\partial {\cal T}_{ij}}{\partial \alpha_{i'}} = 0,\ \ \
\ \frac{\partial \kappa_j^{(n)}(h)}{\partial t_n} = 0.
$$
One can also
introduce $s_{A_j} = \oint_{A_j}ds = \alpha_j$. These are of course
integrals of $t$-flows, in the sense that
 $ \frac{\partial s_{A_j}}{\partial t_n} = 0$, while
$\frac{\partial s_{A_j}}{\partial \alpha_i} = \oint_{A_j} d\omega_i
= \delta_{ij}$. Exponent $f(\alpha_i,t_n|h)$ in (\ref{KPtau})
is certain second-order polinomial in $\alpha_i$ and $t_n$,
though a non-trivial function of the moduli $h$.

Given (i) and (ii), the frequencies $\kappa^{(n)}_j$ can be also
represented as residues:\footnote{
This follows from the identity $d\Omega_n\wedge d\omega_j = 0$
after integration over entire Riemann surface ${\cal C}$:
$$
0 = \int_{{\cal C}}  d\Omega_n\wedge d\omega_j =
\int_{{\cal C}} d(\Omega_nd\omega_j) =
\sum_{i=1}^p \left(\oint_{A_i}d\Omega_n\oint_{B_i}d\omega_j -
\oint_{B_i}d\Omega_n\oint_{A_i}d\omega_j\right)
+ 2\pi i\cdot{\rm res}_{\xi_\pm}\left[ \Omega_nd\omega_j\right],
$$
together with the asymptotical convention (ii).
\label{resref}
}
\be
\kappa^{(\pm n)}_j = \oint_{B_j}d\Omega_{\pm n} =
\frac{1}{n}\oint_{\xi_\pm} (\xi - \xi_\pm)^{-n}d\omega_j(\xi) =
\frac{2\pi i}{n}{\rm res}_{\xi_\pm}(\xi - \xi_\pm)^{-n}d\omega_j,
\label{kappares}
\ee
so that
$$
s_{B_j} = \oint_{B_j} \left(\sum_{i=1}^p \alpha_id\omega_i\right) +
2\pi i\cdot{\rm res}\left(\sum_{n>0} \frac{t_n}{n}(\xi - \xi_+)^{-n} +
\sum_{n>0} \frac{t_{-n}}{n}(\xi - \xi_-)^{-n} +
t_0\log\frac{\xi - \xi_+}{\xi - \xi_-}\right)d\omega_j.
$$

\subsection{Relation between bases}

The basis elements  $d\hat\Omega_n$ and $d\Omega_n$ are related, the
difference between them being somehow expressed through
holomorphic differentials (just because (i) and (ii) define $d\Omega_n$
modulo any combination of $d\omega_i$ - which is then fixed in
a different ways either by (\ref{dSdef'}) or by (iii)):
\be
d\hat\Omega_n \cong d\Omega_n + \sum_{i=1}^p c_i^{(n)}d\omega_i.
\label{OhatO}
\ee
Clearly, coefficients are given by
\be
c_i^{(n)}(h) = \oint_{A_i}d\hat\Omega_n.
\label{cfroOhat}
\ee

\sect{Whitham flows \label{Whsec}}

\subsection{Introducing $a$, $T$ and $d{\cal S}$}

As a first step towards construction of a
 ''quasiclassical $\tau$-function'' -
of which the SW prepotential is an example - let us introduce a solution
 $d{\cal S}(T_n|h_k)$ of  (\ref{dSdef}), depending on (infinitely many)
''slow''  time-variables $T_n$, such that
\be
\frac{\partial \ d{\cal S}}{\partial  T_n} \cong d\Omega_n.
\label{dcalSdef}
\ee
Essentially,   $d{\cal S}(T_n|h_k)$ is a kind of generating function
for all the solutions of (\ref{dSdef}).
Nowadays, the special notation is used for
the periods ${\cal S}_C \equiv \oint_C d{\cal S}$
(motivated by relation to the $d=4$ gauge theory):
\be
a_j \equiv \oint_{A_j}d{\cal S},  \ \ \ \ a^D_j \equiv \oint_{B_j} d{\cal S},
\label{adef}
\ee
and we have:
\be
\frac{\partial a_j}{\partial T_n} = 0, \ \ \ \
\frac{\partial a^D_j}{\partial T_n} = \kappa_j^{(n)}(h).
\label{a/T}
\ee

\subsection{Whitham equations for $h(T)$}

Being solution to (\ref{dSdef}), $d{\cal S}$ can be expanded in the
linear basis $\left\{d\hat\Omega_n\right\}$:
\be
d{\cal S} \cong \sum_{m=-\infty}^\infty u_m(T)d\hat\Omega_m(h),
\label{SinOhat}
\ee
where coefficients $u_m(T)$ are independent of $h$.
Substituting (\ref{SinOhat}) into (\ref{dcalSdef}), we obtain:
\be
\frac{\partial\ d{\cal S}}{\partial T_n} \cong
\sum_{m=-\infty}^\infty \left(\frac{\partial u_m}{\partial T_n}
d\hat\Omega_m \ + \ u_m\frac{\partial\ d\hat\Omega_m}{\partial h_k}
\frac{\partial h_k}{\partial T_n}\right) \
\stackrel{(\ref{dSdef'})}{\cong}\  \nn \\
\cong \sum_{m=-\infty}^\infty \left(\frac{\partial u_m}{\partial T_n}
d\hat\Omega_m \ + \ u_m
\sum_{k} \frac{\partial h_k}{\partial T_n}
\sum_{i=1}^p \sigma_{ki}^{(m)}(h)d\omega_i\right).
\ee
On the other hand,
$ \frac{\partial\ d{\cal S}}{\partial T_n} \cong
d\Omega_n \cong d\hat\Omega_n -
\sum_{i=1}^pc_i^{(n)} d\omega_i$, and from comparison of these
expressions we derive:
\be
\frac{\partial u_m}{\partial T_n} = \delta_{mn}, \ \ {\rm i.e.} \ \
u_m(T) = T_m,
\label{u=T}
\ee
along with the ''Whitham equations'',
\be
\sum_{k}\frac{\partial h_k}{\partial T_n}
\left(\sum_{m=-\infty}^\infty T_m\sigma_{ki}^{(m)}\right)  \equiv
\sum_{k}\frac{\partial h_k}{\partial T_n}\sigma_{ki} = -c_i^{(n)}.
\label{Wheq}
\ee
Condensed notation
$\sigma_{ki} \equiv \sum_{m=-\infty}^\infty T_m\sigma_{ki}^{(m)}$
will be of  use in the further manipulations.

Thus we see that the moduli $h_k$ that were integrals of $t$-flows,
are unavoidably $T$-dependent.

\subsection{Canonical form of $d{\cal S}$}

Equation (\ref{u=T}) implies now that
\be
d{\cal S} \cong \sum_{m=-\infty}^\infty T_md\hat\Omega_m(h|T)
\ \stackrel{(\ref{OhatO})}{\cong}\ \sum_{m=-\infty}^\infty
\left(T_md\Omega_m + T_m\sum_{i=1}^pc_i^{(m)}d\omega_i\right).
\label{dS1}
\ee
Integrating this along $A_j$-cycles and using (\ref{cfroOhat}) we get:
\be
a_j = \oint_{A_j}d{\cal S} =
\sum_{m=-\infty}^\infty T_mc_j^{(m)} =
\sum_{m=-\infty}^\infty T_m\oint_{A_j} d\hat\Omega_m,
\label{afroOhat}
\ee
and substituting this back into (\ref{dS1}),
\be
d{\cal S} \cong \sum_{i=1}^\infty a_id\omega_i +
\sum_{m=-\infty}^\infty T_md\Omega_m.
\ee

Note that,  despite $a_i = a_i(h(T))$ depend on  $h_k$,
which are $T$-dependent,
this dependence cancels completely in $a_i(h_k)$ -
in accordance with (\ref{a/T}).
One can say also that solution to Whitham differential equations
(\ref{Wheq}) contain integration constants - and these are
precisely the $a_i(h)$. This implies that one can add $a_i$ as
extra independent variables to the set of ''slow'' times $T$, and
$d{\cal S} = d{\cal S}(a_i, T_n)$.

Once this is done,  (\ref{dcalSdef}) should be supplemented by an
expression for  $\frac{\partial  d{\cal S}}{\partial a_i}$. However,
it can not be postulated arbitrarily, since we already defined
$d{\cal S}$ unambigously: this expression should be {\it derived}.
It is at this point that the restriction $K = p$ is used for the first time.
Let us consider a variation of (\ref{afroOhat}):
\be
\delta a_j = \sum_{m=-\infty}^\infty \left(\delta T_m\oint_{A_j}
d\hat\Omega_m +
T_m \sum_k \delta h_k\oint_{A_j}\frac{\partial\ d\hat\Omega_m}
{\partial h_k} \right) =
\sum_k \delta h_k \sigma_{kj} + \sum_{m=-\infty}^\infty\delta T_m c_j^{(m)}.
\label{deltaa}
\ee
When $\#k= K = p = \#i$, $\ \sigma_{kj}$ is a square matrix and can be
inverted. Thus, puting $\delta T_m = 0$ in (\ref{deltaa}), we get:
\be
\left. \frac{\partial h_k}{\partial a_j}\right|_{T_n = {\rm const}} =
\sigma^{-1}_{jk}(h).
\label{hvera}
\ee
Consequently,
\be
\left.\frac{\partial\ d{\cal S}}{\partial a_j}\right|_{T_n = {\rm const}}
\cong \sum_{m=-\infty}^\infty T_m\frac{\partial d\hat\Omega_m}{\partial a_j}
\cong \sum_{m=-\infty}^\infty T_m\sum_k \frac{\partial h_k}{\partial a_j}
\frac{\partial d\hat\Omega_m}{\partial h_k} \cong \nn \\ \cong
\sum_k \sigma_{jk}^{-1}\sum_{m=-\infty}^\infty T_m\sigma_{ki}^{(m)}
d\omega_i = \sum_k \sigma_{jk}^{-1}\sigma_{ki}d\omega_i = d\omega_j.
\nn
\ee

To summarize,
\be
d{\cal S} \cong \sum_{m=-\infty}^\infty T_md\hat\Omega_m \cong
\sum_{i=1}^pa_id\omega_i + \sum_{m=-\infty}^\infty T_md\Omega_m,
\nn \\
\frac{\partial  d{\cal S}}{\partial a_i}\ \stackrel{K = p}{\cong}
d\omega_i, \ \ \ \ \
\frac{\partial d{\cal S}}{\partial  T_n} \cong d\Omega_n.
\label{dcalSprop}
\ee
We emphasize that these simple formulas hold, despite the fact that
$a_i$, $d\omega_i$ and $d\Omega_n$ depend (only) on the moduli
$h_k$, which in turn depend non-trivially on $T$'s: $h_k(T)$.
Moreover, after $a_i$'s are introduced as independent variables,
and whenever eq.(\ref{hvera}) can make sense (e.g. if $K = p$),
we rather write $h_k(a_i,T_n)$ instead of $a_i(h_k(T))$.

\subsection{Homogeneity of moduli}

The functions $h_k(a_i,T_n)$, though quite complicated, are
always homogeneous:
\be
\left(\sum_{i=1}^p a_i\frac{\partial}{\partial a_i} +
\sum_{n = 0}^\infty T_n\frac{\partial}{\partial T_n}\right)h_k = 0.
\label{homh}
\ee
This is a simple corollary of (\ref{deltaa}), which can be
rewritten as
$$
\delta h_k = \sum_{j=1}^p \left(\delta a_j -
\sum_{m=-\infty}^\infty c_j^{(m)}\delta T_m\right)\sigma_{jk}^{-1}.
$$
For specific (scale) variation
$\delta a_j = \varepsilon a_j$, $\ \delta T_n = \varepsilon T_n$,
we get:
$$
\delta_\varepsilon h_k = \varepsilon\sum_{j=1}^{p}
\left(a_j - \sum_{m=-\infty}^\infty c_j^{(m)}T_m\right)\sigma_{jk}^{-1}
\ \stackrel{(\ref{afroOhat})}{=} 0,
$$
as stated in (\ref{homh}). This homogeneity condition is of crucial importance
for the possibility to introduce prepotential in any explicit form.

\section{Prepotential. The standard case \label{pre}}
\setcounter{equation}{0}

\subsection{Motivation \label{motiv}}

Let us turn now to the $B$-periods of $d{\cal S}$:
\be
a_j^D = \oint_{B_j}d{\cal S} =
\oint_{B_j}\left(\sum_{i=1}^p a_id\omega_i +
\sum_{n=-\infty}^\infty T_n d\Omega_n\right) =
\sum_{i=1}^p a_i{\cal T}_{ij}(h) + \sum_{n=-\infty}^\infty
T_n\kappa_j^{(n)}(h).
\ee
Further, evaluate
\be
\frac{\partial a_j^D}{\partial a_i} =
\oint_{B_j}\frac{\partial\ d{\cal S}}{\partial a_i} =
\oint_{B_j}d\omega_i = {\cal T}_{ij}(h).
\label{aD/a}
\ee
Since the period matrix is symmetric, ${\cal T}_{ij} = {\cal T}_{ji}$,
this implies that
\be
a_j^D = \frac{\partial {\cal F}}{\partial a_j}
\label{aDfroF}
\ee
for some function ${\cal F}(a_i, T_n)$ often called a
(logarithm of a)
''quasiclassical $\tau$-function'' or, simply, a prepotential.
The arguments of this function are essentially the moduli
of {\it solutions} to original (KP/Toda-like or low-energy Yang-Mills)
dynamical system,
with $a_i(h_k)$ parametrizing the moduli of the curve
${\cal C}$ and $T_n$ - those of coordinate systems in the
vicinity of the punctures.

If ${\cal F}$ was non-singular when {\it all} the times $T_n = 0$,
one could use the homogeneity property (\ref{homh}) to
write down the ''reduced'' prepotential $\hat{\cal F}_{\rm red}(a_i)
\equiv \left.\hat{\cal F}(a_i,T_n)\right|_{{\rm all}\ T_n = 0}$ in a
rather explicit form:
\be
\hat{\cal F}_{\rm red}(a_i) = \frac{1}{2}\sum_{i,j=1}^p a_ia_j
{\cal T}_{ij}(h).
\ee
For a proof, note first of all that this is not a trivial formula,
since $h$ depend non-triviall on $a_i$'s. However, $h_k(a_i)$
are homogeneous of degree zero, thus the same is true about
${\cal T}_{ij}(h(a))$, and consequently $\hat{\cal F}_{\rm red}$
is homogeneous of degree 2:
$$
\hat{\cal F}_{\rm red} = \frac{1}{2}\sum_{i=1}^p a_i
\frac{\partial \hat{\cal F}_{\rm red}}{\partial a_i} =
$$
$$
= \frac{1}{2}\sum_{i=1}^p a_ia^D_i \stackrel{(\ref{adef})}{=}
\frac{1}{2}\sum_{i=1}^p a_i\oint_{B_i}d{\cal S}
\stackrel{(\ref{dcalSprop})}{=} \frac{1}{2}\sum_{i,j=1}^p
a_ia_j{\cal T}_{ij}(h).
$$
Unfortunately, the point $\{all\ T_n = 0\}$ is usually singular,\footnote{
The easiest way to see the singularity is  to note that the matrix
$\sigma_{ki} \equiv \sum_{m=-\infty}^\infty T_m\sigma^{(m)}_{ki}$
is vanishing when all $T_n = 0$, thus $\sigma_{ik}^{-1}$ in
(\ref{hvera}) does not exist. It is also clear that whenever
$\# k = p = \#i$ in (\ref{hvera}) there can be no extra constraints
(like the homogeneity condition) for the functions $h_k(a_i)$ to
satisfy: they describe a {\it non-degenerate} change of
variables. It may be enough, however, to introduce just {\it one} extra
variable (like $T_0$, $T_1$, or - differently - $h_0$) to make
(\ref{hvera}) consistent with homogeneity constraint - and thus allow
some simple truncation of ${\cal F}$ to exist. This is the trick, often
used in considerations of Seiberg-Witten theory: see \cite{EY} as well
as s.\ref{minsec} and s.\ref{Exa} below.
}
$\hat{\cal F}_{\rm red}$ is not defined and this simple reasoning
does not work: the $T$-dependence of ${\cal F}$ should be taken
into account.

Though we did not yet define $\frac{\partial{\cal F}}{\partial T_n}$
explicitly, there is not too much freedom left. Indeed, from
eq.(\ref{aDfroF})
\be
\frac{\partial}{\partial a_i}
\left(\frac{\partial{\cal F}}{\partial T_{\pm n}}\right)
= \frac{\partial^2{\cal F}}{\partial a_i\partial T_{\pm n}}
\stackrel{(\ref{aDfroF})}{=} \frac{\partial a_i^D}{\partial T_{\pm n}}
\stackrel{(\ref{a/T})}{=} \kappa_i^{(\pm n)}
\stackrel{(\ref{kappares})}{=} \frac{1}{n}\oint_{\xi_\pm}
(\xi - \xi_\pm)^{-n}d\omega_i(\xi).
\label{secderF}
\ee
Since from (\ref{dcalSprop}) $d\omega_i \cong \frac{\partial{\cal S}}
{\partial a_i}$, we conclude that
\be
\frac{\partial {\cal F}}{\partial T_{\pm n}} = \frac{1}{n}
\oint_{\xi_\pm}(\xi - \xi_\pm)^{-n} d{\cal S}(\xi), \ \ \ n \geq 1, \nn \\
\frac{\partial {\cal F}}{\partial T_0} =
\oint_{{\rm around\ the\ cut}} \log\frac{\xi - \xi_+}{\xi - \xi_-}
d{\cal S}(\xi) = 2\pi\int_{\xi_-}^{\xi_+}d{\cal S}(\xi).
\label{FoverT}
\ee
This should be added to (\ref{aDfroF}),
\be
\frac{\partial{\cal F}}{\partial a_i} = \oint_{B_i}d{\cal S}.
\label{Fovera}
\ee

In fact, our reasoning after (\ref{secderF}) define $\frac{\partial{\cal F}}
{\partial T_n}$ up to integration ''constants'', which can depend only
on $T$'s, but not on $a$'s. Actually, ${\cal F}$ is
defined modulo addition of any {\it quadratic} function of $T$'s,
$\sum_{m,n=-\infty}^\infty f_{mn}T_mT_n$ with constant coefficients
$f_{mn}$. In practice this is never important, except for one place:
the integral for $\frac{\partial{\cal F}}{\partial T_0}$ in (\ref{FoverT})
is actually divergent, since $d{\cal S}(\xi)$ has poles at $\xi_\pm$.
However, there was no singularity of this kind in (\ref{secderF}):
for $n=0$ it  was just
$\oint_{{\rm cut}} \log\frac{\xi - \xi_+}{\xi - \xi_-}
d\omega_i(\xi) = 2\pi\int_{\xi_-}^{\xi_+} d\omega_i$.
Accordingly, the divergent
part of $ \frac{\partial{\cal F}}{\partial T_0}$
can be absorbed in allowed redefinition of ${\cal F}$.
\subsection{Homogeneity of ${\cal F}$}

Homogeneity property (\ref{homh}) implies that the prepotential
${\cal F}(a_i,T_n)$ always is a homogeneous function of degree 2:
\be
{\cal F} = \frac{1}{2}\left(\sum_{i=1}^p a_i\frac{\partial{\cal F}}{\partial
a_i}
+ \sum_{n=-\infty}^\infty T_n\frac{\partial{\cal F}}{\partial T_n}\right).
\label{homF}
\ee
Indeed, {\it assuming} that (\ref{homF}) is true, and making use of
(\ref{Fovera}) and  (\ref{FoverT}),
\be
{\cal F} = \frac{1}{2}\sum_{i=1}^p a_i\oint_{B_i}d{\cal S} +
\frac{1}{2} \sum_{n>0} \frac{2\pi i}{n}T_{\pm n}{\rm res}_{\xi_\pm}
(\xi - \xi_\pm)^{-n}d{\cal S}(\xi)  + \pi T_0 \int_{\xi_-}^{\xi_+}
d{\cal S}
\stackrel{(\ref{dcalSprop})}{=} \nn \\
= \frac{1}{2}\sum_{i,j=1}^p a_ia_j{\cal T}_{ij}(h) +
\sum_{i=1}^p\sum_{n=-\infty}^\infty a_iT_n\kappa_i^{(n)}(h) +
\frac{1}{2}\sum_{m,n} \frac{2\pi i}{n} T_{\pm n}T_m{\rm res}_{\xi_\pm}
\left[(\xi - \xi_\pm)^{-n}d\Omega_m(\xi)\right].
\label{explF}
\ee
All the coefficients at the r.h.s. (including the residue in the last term)
are functions of moduli $h$ only, thus - given (\ref{homh}) -
what we obtain is indeed homogeneous of degree 2 and our assumption
(\ref{homF}) is {\it a posteriori} justified.

''Explicit formula'' (\ref{explF}) should be of course supplemented
by relations (\ref{hvera}) and  (\ref{Wheq}):
\be
\frac{\partial h_k}{\partial a_j} = \sigma_{jk}^{-1}(h), \ \ \ \
\frac{\partial h_k}{\partial T_n} = \sum_{i=1}^p
 c_i^{(n)}(h)\sigma_{ik}^{-1}(h).
\label{Wheq'}
\ee
 From (\ref{dcalSprop}) one can also obtain:
\be
a_i = \oint_{A_i}d{\cal S}, \ \ \ \
T_{\pm n} = \frac{1}{2\pi i}\oint(\xi - \xi_\pm)^{+n}d{\cal S}
\label{atfrodS}
\ee
(in the second case the property (ii) of $d\Omega$ is used).
Substituting (\ref{FoverT}), (\ref{Fovera}) and (\ref{atfrodS})
 back into (\ref{homF}), we get:
\be
{\cal F} = \frac{1}{2} \sum_{i=1}^p \oint_{A_i}d{\cal S}\oint_{B_i}d{\cal S}
+ \sum_{n >0}
\frac{1}{4\pi in}\oint_{\xi_\pm}(\xi - \xi_\pm)^{+n}d{\cal S}
\oint_{\xi_\pm} \frac{d{\cal S}}{(\xi - \xi_\pm)^{n}} \ + \
\frac{1}{2}\oint_{\xi_+}d{\cal S}\int_{\xi_-}^{\xi_+}d{\cal S}   = \nn \\
= \frac{1}{2} \sum_{i=1}^p \oint_{A_i}d{\cal S}\oint_{B_i}d{\cal S}
+ \frac{1}{2\pi i}\int\int
\log(\xi - \xi')d{\cal S}(\xi)d{\cal S}(\xi').
\label{bilF}
\ee
Note, that while $\frac{\partial {\cal F}}{\partial T_n}$ in (\ref{FoverT})
and $T_n$ in (\ref{atfrodS}) depend on the choice of coordinate
system (e.g. change under rescaling $\xi \rightarrow {\rm const}\cdot\xi$),
this dependence cancels in (\ref{bilF}).

\subsection{Discussion}

Eq.(\ref{bilF})  is a kind of analogue of Hirota bilinear equation for KP/Toda
$\tau$-function, and in this analogy the $1$-differential $d{\cal S}(\xi)$
substitutes the Baker-Akhiezer $\frac{1}{2}$-differential $\Psi(\xi)$.
Note that (\ref{bilF}) depends on the data $\{$the curve ${\cal C}$,
points $\xi_\pm$, coordinate system$\}$ - which is exactly the
data, specifying particular {\it solution} to Hirota equations.
This reflects the fact that the set of Whitham {\it equations} depends
on both, original system (e.g. KP/Toda hierarchy) and its solution.
The main drawback of (\ref{bilF}) in its present form is the lack of
alternative expression of $d{\cal S}(\xi)$ through ${\cal F}$
- while for the usual KP/Toda Baker-Akhiezer function
\be
\Psi(\xi) = \frac{\tau\left(t_{\pm n} - \frac{1}{n}(\xi - \xi_\pm)^n\right)}
{\tau(t_{\pm n})}\exp \left(\sum_n t_{\pm n}(\xi - \xi_\pm)^{-n}\right).
\label{BAf}
\ee
This is what makes the theory of prepotential with infinitely
many time-variables - at the present stage -
dependent on the formalism of KP/Toda-hierarchies, involving
spectral {\it curves} (not  hypersurfaces), 1-differentials etc - and makes
applications beyond such framework not straightforward.
Such applications, of course, exist: in the context of Seiberg-\-Witten
theory the simplest example is its reformulation in terms of Calabi-Yau
manifolds \cite{CY,KKLMV,MW2}.

Moreover, even in the case of  spectral {\it curves}, above consideration
is not exhaustive, because it relies heavily on 1-differentials, rather than
on 2-differentials, which are better suited for consideration of generic
moduli spaces of complex curves. In fact, the whole theory of prepotential
is the one about cohomology groups, more exactly Hodge structures
over the moduli spaces and is often formulated this way in considerations
starting from topological field theories, see \cite{D,Kr2} and
 references therein.

\subsection{Homogeneity property in terms of $d{\cal S}$
\label{minsec}}

Eq.(\ref{homF}) can be reexpressed in terms of $d{\cal S}$,
what is useful both for applications and for generalizations.
For this, differentiate (\ref{homF}) w.r.to $h_k$,
keeping all the $T_n$ fixed:
\be
\frac{\partial}{\partial h_k}\left( 2{\cal F} - \sum_{i=1}^p
a_i\frac{\partial F}{\partial a_i}\right) =
\sum_{i=1}^p \left(\frac{\partial a_i}{\partial h_k}
\frac{\partial{\cal F}}{\partial a_i} - a_i
\frac{\partial}{\partial h_k}\frac{\partial{\cal F}}{\partial a_i}\right)
= \sum_{n=-\infty}^{+\infty} T_n\frac{\partial}{\partial h_k}
\frac{\partial{\cal F}}{\partial T_n}
\label{App1}
\ee
Using (\ref{aDfroF}), the l.h.s. can be also rewritten as
\be
\sum_{i=1}^p \left(\frac{\partial a_i}{\partial h_k} a_i^D -
a_i\frac{\partial a_i^D}{\partial h_k}\right) =
\sum_{i=1}^p\left(\oint_{A_i}\frac{\partial d{\cal S}}{\partial h_k}
\oint_{B_i}d{\cal S} - \oint_{B_i}\frac{\partial d{\cal S}}{\partial h_k}
\oint_{A_i}d{\cal S}\right) =
\oint_{{\rm sing}} {\cal S}\frac{\partial d{\cal S}}
{\partial h_k}.
\label{App2}
\ee
where the integral at the r.h.s. runs around the singularities
of $d{\cal S}$ and $\partial d{\cal S}/\partial h_k$.
The last identity in (\ref{App2})
is valid for {\it any} (1,0)-differential $d{\cal S}$
(it results from integration of identity
$d{\cal S}\wedge\frac{\partial d{\cal S}}{\partial h_k} = 0$
over entire Riemann surface ${\cal C}$, comp.with footnote
\ref{resref}). Assuming that $d{\cal S}$ is meromorphic with poles
only at $\xi_\pm$ and $\frac{\partial d{\cal S}}{\partial h_k}$
is holomorphic, one can
rewrite the r.h.s. of (\ref{App2}) in a more explicit form:
\be
{\rm l.h.s.of}\ (\ref{App1}) \stackrel{(\ref{aDfroF})}{=}
\oint_{{\rm sing}} {\cal S}
\frac{\partial d{\cal S}}{\partial h_k} = \nn \\
= \frac{1}{i}\oint_{\xi_+}d{\cal S}(\xi)
\int_{\xi_-}^{\xi_+}\frac{\partial d{\cal S}(\xi')}{\partial h_k} \ +
\sum_{n=1}^\infty \frac{1}{2\pi in}\left(
\oint_{\xi_+}(\xi - \xi_+)^n d{\cal S}(\xi)
\oint_{\xi_+}(\xi' - \xi_+)^{-n}\frac{\partial d{\cal S}(\xi')}{\partial h_k} +
\right. \nn \\ \left.
+ \oint_{\xi_-}(\xi - \xi_-)^n d{\cal S}(\xi)
\oint_{\xi_-}(\xi' - \xi_-)^{-n}\frac{\partial d{\cal S}(\xi')}{\partial h_k}
\right).
\label{App3}
\ee
It is easy to recognize in this expression the r.h.s. of (\ref{App1})
with (\ref{FoverT}) and (\ref{atfrodS}) substituted in it.

Holomorphicity of $\frac{\partial d{\cal S}}{\partial h_k}$, required
in this derivation, is exactly the requirement (\ref{dSdefor}),
which is in fact the crucial reason for (\ref{aD/a}) and (\ref{aDfroF})
to hold. Thus, as we see once again, from certain point of view
the $T$-dependence does not introduce anything essentially new:
given $d{S}$ that satisfies (\ref{dSdef}) and thus (\ref{aDfroF}),
eqs.(\ref{FoverT}) and (\ref{atfrodS}) can be just used to {\it define}
what are $T_n$ and $\frac{\partial {\cal F}}{\partial T_n}$.

Moreover, the holomorphicity requirement  (\ref{dSdefor})
implies that so defined $T_{\pm n} = \oint_{\xi_\pm}
(\xi - \xi_\pm)^nd{S}(\xi)$ are independent of $h_k$ (because
$\ {\rm res}_{\xi_\pm}(\xi - \xi_\pm)^{n} d\omega_j(\xi) = 0$
for any $n\geq 0$) - and thus (\ref{App3}) can be explicitly integrated
over $h_k$: it is enough to substitute
$\frac{\partial d{\cal S}(\xi')}{\partial h_k} \rightarrow  d{\cal S}(\xi')$.
In other words, we obtain:
\be
2{\cal F} - \sum_{i=1}^pa_i\frac{\partial {\cal F}}{\partial a_i} = \nn \\
= \frac{1}{ i}\oint_{\xi_+} d{\cal S} \int_{\xi_-}^{\xi_+} d{\cal S}
+ \sum_{n =1}^\infty \frac{1}{2\pi in}\sum_{\alpha = \pm}
\oint_{\xi_\alpha} (\xi - \xi_\alpha)^nd{\cal S}(\xi)
\oint_{\xi_\alpha} (\xi' - \xi_\alpha)^{-n}d{\cal S}(\xi').
\label{homFh}
\ee
Once a particular solution $dS(h_k)$ of (\ref{dSdefor}) is given,
the r.h.s. of (\ref{homFh}) is explicit function of $h_k$.

Of course, concrete solution corresponds to concrete values of
time-variables $T_n$ - and in this sense eq.(\ref{homFh}) is
somewhat artificial. However, according to \cite{EY},
such approach, based on (\ref{bilF}) and (\ref{homF})  is instead
useful for comparison of the abstract prepotential theory with
numerous particular results, obtained by more sophisticated
methods \cite{Pic,STY}, see also s.\ref{Exa} below.
Moreover it can be immediately
generalized beyond spectral {\it curves} - to (certain) complex
manifolds of arbitrary dimension.

\sect{Prepotential beyond one complex dimension \label{genpre}}

We can now reverse the logic and consider eq.(\ref{bilF})
as the starting point {\it instead} of eq.(\ref{dSdefor}). The
disadvantage of such approach is that we can loose understanding
of what is original  model,  to which our consideration is a
Whitham approximation - in the case of (\ref{dSdef}) it was
a system of KP/Toda family  (and - less obviously - the $N=2$
gauge models in $4d$). Of course, this is also the main
advantage: this opens the way to considerations in the
spirit of integrability theory beyond the KP/Toda framework.
As we shall see, there is no problem of developing the literal
analogue of above Whitham theory for certain complex
manifolds of any dimension instead of complex curves (with
immediate examples provided by K3 and Calabi-Yau manifolds).
The analogue of entire KP/Toda theory for these examples
should be intrinsically stringy and probably include multi-loop
groups, but the Whitham approximation is very similar to KP/Toda
case - as one expects a kind of simplified universal description
emerges in the ''slow'' or ''low-energy'' limit. With no
surprise this higher-dimensional extension of Whitham formalism
is very close to ''special geometry'' \cite{SG}, as well as to
the abstract constructions of \cite{Kr2} and \cite{D}.

The content of this section is very much a repetition of what was
said in s.\ref{pre}: but now the accents and the whole logic are
different - and just this rephrasing makes possible a
much wider applications of  essentially the same formulas.

\subsection{Notations and definition}

Consider a family ${\cal M}(h)$
of complex  manifolds $M$ of complex dimension $d$ (in the
previous sections $d=1$ and ${\cal M}(h)$ are some families
of spectral {\it curves}, $M = {\cal C}$ ).
 The family is para\-met\-ri\-zed by some {\it moduli} $h_k$,
$k = 1,\ldots,K = {\rm dim}_C{\cal M}$.
Let us fix some canonical system  of $d$-cycles on $M$:
$\ \{A_i, B_i\}$, $i = 1,\ldots,p = \frac{1}{2}{\rm dim} H^{d}(M)$
with the intersection matrix $A_i\#B_j = \delta_{ij}$,
$A_i\#A_j = B_i\#B_j = 0$. Finaly, pick up some holomorphic
$(d,0)$-form $\Omega$ on every $M$.\footnote{
It is clearly
a restriction on $M$ that such $\Omega$ exists,
examples of suitable $M$ are provided by $K3$ ($d=2$) and
Calabi-Yau ($d=3$) manifolds.
In our discussion below we shall see that this restriction
can sometime be weekend, by admitting $\Omega$'s
with simple singularities.
Additional requirements
for $\Omega$-dependence on moduli -
in the spirit of (\ref{dSdefor}) - will be specified later.
}
Its {\it periods},
\be
a_i(h) \equiv \oint_{A_i}\Omega, \ \ \ \
a^D_i(h) \equiv \oint_{B_i}\Omega
\ee
are functions of moduli.

Consider now a variation $\delta\Omega$ of $\Omega$ with
the change of parameters (moduli). $\delta\Omega$ is also a
$(d,0)$-form, not necessarily holomorphic.
Still, always $\Omega\wedge\delta\Omega = 0$
(just because $\Omega$ is a maximal-rank form), and integraion of
this relation over entire $M$ gives:
\be
0 = \int_M\Omega\wedge\delta\Omega =
\sum_i\left(\oint_{A_i}\Omega\oint_{B_i}\delta\Omega -
\oint_{A_i}\delta\Omega\oint_{B_i}\Omega\right) \ + \
{\rm contribution\ from\ singularities}.
\label{origin}
\ee
Imagine that the last item at the r.h.s. - the contribution from
singularities of $\Omega$ and $\delta\Omega$ is absent.
Then we obtain from (\ref{origin}):
\be
\sum_i a_i\delta a^D_i = \sum a^D_i\delta a_i.
\label{symrel}
\ee
This implies that the {\it prepotential}, defined as -
\be
{\cal F} \equiv \frac{1}{2}\sum_i a_ia^D_i =
\frac{1}{2}\sum_i \oint_{A_i}\Omega\oint_{B_i}\Omega,
\label{Fdef}
\ee
possesses the following property:
\be
\delta{\cal F} = \frac{1}{2}\sum_i\left(a_i\delta a^D_i + a^D_i\delta
a_i\right)
= \sum_i a^D_i\delta a_i.
\ee
If the freedom of variations is big enough, e.g. if
$\#K = {\rm dim}_C{\cal M}$
is the same as $\#p = \frac{1}{2}{\rm dim}\ H^{d}(M)$,
we conclude from this that
\be
a^D_i = \frac{\partial{\cal F}}{\partial a_i}
\label{aDfroF"}
\ee
and
\be
{\cal F} = \frac{1}{2}\sum_i a_ia^D_i =
\frac{1}{2}\sum_i a_i\frac{\partial {\cal F}}{\partial a_i}.
\label{homF"}
\ee
In other words, we can consider $a_i$ as independent variables,
and introduce the prepotential ${\cal F}(a)$ by the rule (\ref{Fdef}) -
and it will always be a homogeneous function of degree 2 -
as follows from (\ref{homF"}).

The two  requirements, built into this simple construction, are:

(i) the absence of singularity contributions at the r.h.s. of (\ref{origin});

(ii) the matching between the quantities of moduli and $A$-cycles,
$$
K \equiv {\rm dim}_C {\cal M} =  p \equiv \frac{1}{2}{\rm dim}\ H^{d}(M)
$$.

\subsection{Comments on requirement (i)}

The problem with this restriction is that variation of holomorphic
object w.r.to moduli usually makes it singular - by the very definition
of moduli of complex structure. Thus, even if $\Omega$ is free of
singularities one should expect them to appear in $\delta\Omega$.
The only way out would be to get the newly emerging poles
canceled by zeroes of $\Omega$ - but often the space of holomorphic
$\Omega$'s is too small to allow for adequate adjustement.

Fortunately, requirement (i) can be made less restrictive.
One can allow to consider $\Omega$ which is not holomorphic,
but possesses {\it simple} singularities at isolated divisors. As
a pay for this it is enough to  enlarge the set of $A$-cycles, by adding
the ones, wrapping around the singularity divisors, and also add all
independent $B$-chains, connecting these divisors (such that
$\partial B = div_1 - div_2$).\footnote{
For example, if $M$ is a complex curve ($d=1$), $\Omega$ can be
a meromorphic $(1,0)$-differential with {\it simple} (order one) poles
at some punctures $\xi_\alpha$, $\alpha = 0,1,\ldots,r$. Then one
should add $r$ circles around the points $\xi_1,\ldots,\xi_r$ to the
set of $A$-cycles, and $r$ lines (cuts) connecting $\xi_0$
with $\xi_1,\ldots,\xi_r$ to the set of $B$-contours in eq.(\ref{origin}).
Then the last term at the r.h.s. can be omitted in exchange for enlarging
the sum in the first term.
}
At the same time residues at the simple singularities should be added to
the set of moduli $\{h\}$, thus preserving the status of the
second requirement (ii).

This prescription is still not complete, because the integrals over
newly-added $B$-chains are divergent (because these end
at the singularities of $\Omega$). However, the structure of divergencies
is very simple: if a cut-off is introduced, the cut-off-\-dependent
piece in ${\cal F}$ is exactly quadratic in the new moduli -
and does not depend on the old ones. If one agrees to
define the prepotential - which is generic homogeneous
function of order two - modulo {\it quadratic} functions of
moduli, the problem is resolved.

Thus the real meaning of constraint (i) is that $\delta\Omega$ should
not introduce new singularities as compared to $\Omega$ - so that we
do not need to introduce new cycles, thus new moduli, derivatives
over which would provide new singularities etc etc.
Since now the freedom to choose $\Omega$ is big enough, such
special adjustement is usually available.

The non-simple singularities (higher-order poles at divisors)
should be resolved - i.e. considered as a limit of several simple
ones when the corresponding divisors tend to coincide. The
corresponding $B$-chains shrink to zero in the limit, but integrals
of $\Omega$ over them do not vanish, if $\Omega$ is indeed
singular enough. This procedure of course depends on particular
way to resolve the non-simple singularity. Essentially, if we
want to allow  the one of arbitrary type on the given divisor,
it is necessary to introduce coordinate system in the vicinity
of divisor and consider all the negative terms of Laurent
expansion of $\Omega$ as moduli, and ''weighted'' integrals
around the divisor as $A$-cycles. In the case of $d=1$, when
the divisors are just points (punctures) one can easily
recognize in this picture the definition of KP/Toda-induced
Whitham prepotential in the form of (\ref{bilF}):
with one-parameter set of ''time''-variables
(Laurent expansion coefficients) for every puncture as
additional moduli. As often happens, it is most natural from the
point of view of string theory (integrability theory in this case)
to put all the moduli in a single point (or two), but from the
point of view of algebraic geometry it is better to redistribute
them as simple singularities at infinitely many divisors.

Finally, singularities of $\Omega$ on subspaces of codimension higher
than one do not contribute to eq.(\ref{origin}) at all  - and often
variation w.r.to moduli produces only singularities of such type
as $d>1$.

\subsection{Requirement (ii)}

Thus, what essentially remains is the other requirement (ii) -
the matching condition between the number of moduli and $A$-cycles.
Since the procedures involved in resolution of (i) do not change this
matching (they always add as many new moduli as new $A$-cycles),
this requirement can be analyzed at the very beginning -
before even introducing $\Omega$.

The Seiberg-Witten theory of $4d$
low-energy $N=2$ SUSY gauge theories
provides a lot of examples when such matching takes place.
Most of them have immediate counterparts in the $1d$ integrability
theory \cite{Go,MW1,DW,M,GoM,IM}.


\section{Examples \label{Exa}}
\setcounter{equation}{0}

\subsection{Definitions}

In examples below we first list the curves ${\cal C}$, which are
associated with particular $N=2$ SUSY gauge theories and -
at the same time - with particular integrable systems of particles
(see \cite{IM} for details and references). The ''full'' spectral curves
are naturally defined by algebraic equations
\be
{\cal C}: \ \ \ \ \   \det(t - L(z)) = 0
\label{curve}
\ee
with parameter $z$ being a coordinate on elliptic
''bare spectral curve'' $E(\tau)$ - which in many cases degenerates
into a sphere (in a special double scaling limit as
 $\tau \rightarrow i\infty$).
For every example we consider the
''minimal'' $d{\cal S}$ - a solution to (\ref{dSdefor}),
defined by the formula\footnote{
There are various (essentially equivalent) ways to
characterize $d{\cal S}_{\rm min}$: as a minimal solution
to the constraint (\ref{dSdefor}), as reduced $d{\cal S}$
(\ref{dcalSprop}), as $d^{-1}$ of symplectic form
$dt\wedge d\omega_0$ on the Hitchin variety etc.
}
\be
d{\cal S}_{\rm min} \cong 2td\omega_0 (z)
\stackrel{\tau \rightarrow i\infty}{\longrightarrow} \frac{t}{i\pi}d\log z,
\label{diff}
\ee
where $d\omega_0(z)$ is distinguished canonical holomorphic 1-differential
on $E(\tau)$, which in the scaling limit turns into $\frac{dz}{2\pi iz}$  -
or rather into
$\frac{1}{2\pi i}\left(\frac{d\xi}{\xi - \xi_+} -
 \frac{d\xi}{\xi - \xi_-}\right)$.
This $d{\cal S}_{\rm min}$ is actually our $d{\cal S}(a_i,T_n)$
at the hypersurface  where
all the $T$-variables are vanishing - except for one or two, which are
usually $T_{\pm 1}$ or $T_0$. Given $d{\cal S}_{\rm min}$
and accepting the upside-down logic of section \ref{genpre},
one can use (\ref{bilF}) to {\it define} the reduced prepotential
$$
{\cal F}_{\rm red}(a_i) \equiv \left.{\cal F}(a_i,T_n)\right|_{
\stackrel{T_0\ {\rm or}\ T_{\pm 1} = {\rm const}}{{\rm all\ other}
 \ T_n = 0}},
$$
which is not homogeneous as a function of $a_i$, but instead satisfies
\be
2{\cal F}_{\rm red} - \sum_{i=1}^p a_i\frac{\partial{\cal F}_{\rm red}}
{\partial a_i} =  T_0\frac{\partial{\cal F}_{\rm red}}{\partial T_0}
+ T_1\frac{\partial{\cal F}_{\rm red}}{\partial T_1} + T_{-1}
 \frac{\partial{\cal F}_{\rm red}}{\partial T_{-1}}
\stackrel{(\ref{homFh})}{=} \nn \\ =
{2\pi}\cdot{\rm res}_{\xi_+} \left[d{\cal S}_{\rm min}\right]
 \int_{\xi_-}^{\xi_+}
d{\cal S}_{\rm min}\
+ \sum_{\alpha = \pm}
{\rm res}_{\xi_\alpha}
\left[(\xi - \xi_\alpha)d{\cal S}_{\rm min}(\xi)\right]
\oint_{\xi_\alpha} \frac{d{\cal S}_{\rm min}(\xi)}{\xi - \xi_\alpha}.
\label{homFred}
\ee
This was also the logic accepted in \cite{EY}.

Actually, we shall mostly concentrate on the case of $SL(2)$ group.
Of most interest are relations in the chain
{\it pure gauge} $N=2$ -- {\it pure gauge} $N=4$ --
{\it Calabi-Yau compactification}, which can be revealed
already at the $SL(2)$ level.
As the last example, we discuss the Calabi-Yau manifold,
associated with $WP^{12}_{1,1,2,2,6}$. Related string-induced model
is known \cite{KKLMV}
to reproduce $SU(2)$ $N=2$ Seiberg-\-Witten/sine-\-Gordon
theory in the conifold limit.  Beyond conifold limit
one rather expects it to be related in the same sense to the
$SU(2)$ $N=4$ SW/Calogero model:\footnote{
A related claim is made in a recent paper \cite{GHL}.
}
the prepotentials for the
two - when written in appropriate variables -
should be closely related, if not the same.
Exact result should be looked for
at the level of Picard-Fuchs equations. We demonstrate
that certain similarity indeed exists, but exact equality hardly
takes place.

\subsection{Comments}

As we already mentioned in the previous sections,
(\ref{bilF}) is essentially the same as
(\ref{homFred}):
\be
{\cal F}_{\rm red} (a_i) = \frac{1}{2}\sum_{i=1}^p a_ia_i^D +
\pi T_0\int_{\xi_-}^{\xi_+}d{\cal S}_{\rm min} +
\frac{1}{2}\sum_\alpha T_{\alpha 1}
\oint_{\xi_\alpha} \frac{d{\cal S}_{\rm min}(\xi)}{\xi - \xi_\alpha},
\label{Freddef}
\ee
\be
a_i = \oint_{A_i}d{\cal S}_{\rm min}, \ \ \ \
T_0 = {\rm res}_{\xi_+} \left[d{\cal S}_{\rm min}\right] , \ \ \ \
T_1 = {\rm res}_{\xi_\alpha}\left[(\xi - \xi_\alpha)
d{\cal S}_{\rm min}(\xi)\right].
\label{defmod}
\ee
This should be considered as a somewhat {\it im}plicit formula
for ${\cal F}_{\rm red}(a_i)$: eq.(\ref{Freddef}) define it as a function
of $h_k$ (on which $d{\cal S}_{\rm min}$ depends), rather than
that of $a_i$: eq.(\ref{defmod}) should be resolved to give
$h_k(a_i)$ and then the result is to be substituted into (\ref{Freddef}).
It now follows from the general theory above that
\be
\oint_{B_i}d{\cal S}_{\rm min} \equiv a_i^D =
\left.\frac{\partial{\cal F}}{\partial a_i}\right|_{T_0,T{\alpha 1}
 = {\rm const}}.
\ee

In fact, only $a_i$, $a_i^D$ and their dependence on $h_k$
have physical meaning - both in $4d$ and $1d$ models.\footnote{
It deserves reminding that moduli $h_k$ are vacuum expectation values
of certain scalar fields and  eigenvalues of Hamiltonians
in $4d$ and $1d$ respectively. The periods $a_i$ and $a_i^D$
in $4d$ are entering the mass mormula for the BPS-saturated
states, which  are physical (this is significant distinction of $N=2$
SUSY theories - normally the background fields in the Wilsonian
action, like $a_i$ and $a_i^D$, are not physical
observables). In $1d$ the $a_i$ and $a_i^D$ are action integrals.
}
 From this point of view the shapes of the {\it curves} ${\cal C}(h_k)$
are not uniquely defined for the given model. As is well known since
\cite{SW2}, different curves can produce the same periods
$a_i(h)$, $a_i^D(h)$ (with the same $h$-dependence).
Obviously, a {\it covering} of original curve will reproduce the
same periods - but it will be another curve. In eq.(\ref{curve})
this freedom is reflected by that of the choice of representation
of the Lax operator for the given group.

Dependence of $a_i$ and $a_i^D$ on $h_k$ is usefully
encoded in terms of the so-called Picard-Fuchs equations (PFE) -
which immediately arise once integral formulas like
$a_i(h_k) = \oint_{A_i}d{\cal S}_{\rm min}(h_k)$,
$a_i^D(h_k) = \oint_{B_i}d{\cal S}_{\rm min}(h_k)$,
are written down. At least naively, these equations are
{\it additional} pieces of information as compared to the prepotential:
the latter one is a function of $a_i$ and does not contain any
reference to $h_k$. This is also reflected in the fact that details
of differential geometry on particular curves ${\cal C}$ (and
manifolds $M$, in general) are involved in the appearence
of PFE (to say nothing about their derivation):
occurence of the $\sigma_{kj}$ matrix in (\ref{hvera})
is an (oversimplified) example.
Of course, once {\it all} the PFE for some two models
coincide, the same is true about their prepotentials - {\it provided}
they exist: the PFE can be written down for any cohomology
class - not just to the one represented by
$d{\cal S}_{\rm min}(h_k)$.
But if the PFE are written for
{\it this} particular class - they can tell us a lot about the prepotential,
which {\it does} exist because $d{\cal S}_{\rm min}$ satisfies
(\ref{dSdefor}). Similarly, in the framework of special geometry
in higher dimensions - where again the prepotential exists
by definition, - the PFE for cohomology class of $\Omega(h_k)$
can be used for its investigation.

\sect{Examples of UV-infinite $4d$ models: Riemann sphere as the
{\it bare} spectral surface}

As explained in \cite{SW2,DW,IM}, the $d=4$ models which are
asymptotically free in the UV and acquire mass scale as result
of dimensional transmutation, in the language of $1d$
integrable systems correspond to the models with singular
{\it bare} spectral surface. In the simplest setting this can
be just a Riemann sphere with punctures: two for pure
gauge $N=2$ SUSY models and more when extra matter
multiplets are introduced. The ''soft'' UV regularization of the theory
(resulting from its embedding into some UV-finite model)
corresponds to blowing up the singularities (punctures),
making the {\it bare} spectral surface into elliptic curve (torus)
$E(\tau)$: this will be the subject of our consideration in the
s.\ref{Calod} below.

\subsection{Original Seiberg-\-Witten
(sine-\-Gordon) example. Curve and $d{\cal S}_{\rm min}$ \label{oriSW}}

This basic example was  introduced in \cite{SW} for description of
(the low-energy limit of) pure $SU(2)$-gauge $N=2$ SUSY theory
and considered from the point of view of integrability theory
in \cite{Go}: the relevant universality class appears to be
that of $1d$ sine-Gordon model (a particle in cosine potential).
The {\it full} spectral curve
${\cal C}$ is of genus $p=1$ and is defined by elliptic equation
\be
\tilde y^2 = (\tilde x - u)(\tilde x^2 - \Lambda^4),\ \ \ \
u = 2h_2 \ \   (= 2h_2^{N=2} = 2h_2^{{\rm SG}}).
\ee
The degenerated {\it bare} spectral curve is a 1-punctured sphere, and the
associated $d{\cal S}_{\rm min}$ is (see s.\ref{Calod} below for details
of degeneration {\it scaling} limit):
\be
d{\cal S}_{\rm min} \cong \frac{1}{\pi\sqrt{2}}\frac{\tilde x-u}
{\tilde y(\tilde x)}d\tilde x =
\frac{\Lambda}{\pi\sqrt{2}}\sqrt{\frac{u}{\Lambda^2} - \cos\varphi}\ d\varphi =
\frac{1}{2\pi i}
\sqrt{2u + \Lambda^2\left(z+ \frac{1}{z}\right)}\ \frac{dz}{z}.
\label{dSexa1}
\ee
Obviously,
$$
\frac{\partial d{\cal S}_{\rm min}}{\partial u} =
-\frac{1}{2\pi\sqrt{2}}\frac{dx}{\tilde y(x)} = -\frac{1}{2\pi\sqrt{2}}dv(x)
$$
-  in accordance with
(\ref{dSdef}). This $d{\cal S}_{\rm min}$ has double pole at the branch
point $x = \infty$, with vanishing residue. Let us assume  that
coordinate in the vicinity of $x = \infty$ is
$\xi - \xi_\infty = \sqrt{\Lambda^2/x}$.
Then $d{\cal S}_{\rm min} = \frac{\sqrt{2}\Lambda}{\pi} d\hat\Omega_1$,
so that $T_1 = \frac{\sqrt{2}}{\pi}\Lambda$, and all other $T_n = 0$.

\subsection{Prepotential}

According to (\ref{Freddef})
\be
{\cal F}_{\rm red}(a) = \frac{1}{2} aa_D - \frac{iu}{\pi}
\label{Fexasl2}
\ee
Here:
\be
a = \oint_Ad{\cal S}_{\rm min} =
\frac{\sqrt 2}{\pi}\int_{-\Lambda^2}^{+\Lambda^2}
\frac{x-u}{\tilde y(x)}dx =
\frac{\sqrt 2}{\pi}\int_{-\Lambda^2}^{+\Lambda^2}
\sqrt{\frac{x-u}{x^2-\Lambda^4}}\ dx ,
\nn \\
a^D = \oint_B d{\cal S}_{\rm min} = \frac{\sqrt 2}{\pi}
\int_{\Lambda^2}^{u}\sqrt{\frac{x-u}{x^2-\Lambda^4}}\ dx ,
\label{aaDexa}
\ee
and $-\frac{i}{\pi} = \frac{1}{2}\cdot
\left(\frac{\sqrt 2}{\pi}\Lambda\right)^2 \cdot
2\pi i\cdot \left(-\frac{u}{2\Lambda^2}\right)$.
By definition of ${\cal F}_{\rm red}$,
\be
a^D = \frac{\partial{\cal F}_{\rm red}}{\partial a}.
\label{aDfroFexa}
\ee

In the large-$a$ (large-$u$) limit things simplify a lot:
\be
a^D \sim \frac{2i}{\pi}a\log a,   \ \ \ \ \ u \sim \frac{1}{2}a^2,
\label{largeuaD}
\ee
and  (\ref{Fexasl2}) is just the function
\be
{\cal F}_{\rm red}(a) = \int^a a^Dda \sim
\frac{2i}{\pi}\int^a a\log a\ da =
\frac{i}{\pi} a^2\log a - \frac{i}{2\pi}a^2 = \frac{1}{2} aa_D
  - \frac{iu}{\pi}.
\ee

For generic $a$ these formulas are more involved:
to check  (\ref{aDfroFexa}) given (\ref{Fexasl2}) one
should check that the integrals (\ref{aaDexa}) satisfy
\be
a^D = \frac{\partial{\cal F}_{\rm red}}{\partial a} =
\frac{1}{2}a^D + \frac{1}{2} a\frac{\partial a_D}{\partial a} +
\frac{1}{i\pi}\frac{\partial u}{\partial a}
\label{garb1}
\ee
Since
$$
\frac{\partial a}{\partial u} =
- \frac{1}{\pi\sqrt{2}}\int_{-\Lambda^2}^{\Lambda^2}
\frac{dx}{\tilde y(x)} =
- \frac{1}{\pi\sqrt{2}}\int_{-\Lambda^2}^{\Lambda^2}
\frac{dx}{\sqrt{(x-u)(x^2-\Lambda^4)}},
$$
eq.(\ref{garb1}) is an identity for elliptic integrals.
Of course it is immediate corollary of our usual argument
($dv(x) = \frac{dx}{\tilde y(x)}$):
\be
0 = \int_{{\cal C}} d{\cal S}_{\rm min} \wedge dv =
\oint_A d{\cal S}_{\rm min}\oint_B dv - \oint_Bd{\cal S}_{\rm min}
\oint_A dv  - {\rm res}_\infty
\left[ {\cal S}_{\rm min}(\tilde x) dv(\tilde x)\right].
\nn
\ee

The homogeneity equation (\ref{homFred})
for (\ref{Fexasl2}) is just immediate:
\be
2{\cal F}_{\rm red} - a\frac{\partial{\cal F}_{\rm red}}{\partial a}
= (aa^D - \frac{2iu}{\pi}) - aa^D = -\frac{2iu}{\pi}.
\label{homexa1}
\ee

\subsection{Picard-Fuchs equation}

In the case of $SL(2)$, when there is a single module $h_2$,
eq.(\ref{homFh}) - i.e. eq.(\ref{homexa1}) - possesses one more
useful interpretation \cite{Pic}: in terms of Picard-Fuchs
equations.

The starting point is relation of the form
\be
{\cal D}\left(h_k,\frac{\partial}{\partial h_k}\right) \cong 0
\label{Pic1}
\ee
for some differential operator ${\cal D}(h,\partial/\partial h)$,
peculiar for the given moduli space (i.e. the family
of  spectral curves under consideration).
In the particular case of  (\ref{dSexa1}) this identity is
\be
\left(\frac{\partial^2}{\partial u^2} - \frac{1}{4(\Lambda^4 - u^2)}\right)
d{\cal S}_{\rm min} \cong 0.
\label{Pic2}
\ee
Indeed, the l.h.s. is proportional to the total derivative:
\be
0 \cong 2d\sqrt{\frac{\tilde x^2 - \Lambda^4}{\tilde x-u}} =
\frac{\tilde x^2 - 2u\tilde x + \Lambda^4}{\tilde x-u}
\frac{d\tilde x}{\tilde y(\tilde x)} \sim
d{\cal S}_{\rm min} - 4(\Lambda^4 - u^2)\frac{\partial^2}{\partial u^2}
d{\cal S}_{\rm min}.
\nn
\ee

As a relation in cohomologies, (\ref{Pic1}) is essentially an equation
for the periods of $d{\cal S}_{\rm min}$:
$$
{\cal D}\left(\frac{\partial}{\partial h_k}\right) a_i =
{\cal D}\left(\frac{\partial}{\partial h_k}\right) a_i^D = 0.
$$
When there is only one $h_k$
 and ${\cal D}$ is a second order operator
- as is the case for (\ref{Pic2})\footnote{
The reason for this is simple: in such case all the three differentials
$dS$, $\frac{\partial}{\partial h}dS$ and  $\frac{\partial^2}{\partial h^2}dS$
have vanishing residues, and in addition the spectral surface
is of genus one,- therefore their cohomology classes are
distinguished by only {\it two} periods, thus a linear relation
should exist between the three classes. As soon as non-trivial
residues are allowed at $r+1$ points and genus is $p$, the naive
Picard-Fuchs equations should include $r+2p$ derivatives.
} -
\be
{\cal D}\left(\frac{\partial}{\partial h}\right) =
\frac{\partial^2}{\partial h^2} + U(h)\frac{\partial}{\partial h} + V(h),
\label{Pic3}
\ee
this can be transformed into a closed equation for the Wronskian
$$
W \equiv a\frac{\partial a^D}{\partial h} - \frac{\partial a}{\partial h}a^D
= \frac{\partial}{\partial h}\left(2{\cal F} -
a\frac{\partial{\cal F}}{\partial a}\right):
$$
$$
\frac{\partial W}{\partial h} = a\frac{\partial^2 a^D}{\partial h^2}
- \frac{\partial^2a}{\partial h^2}a^D = - U(h) W,
$$
i.s. $W(h) = {\rm const}\cdot \exp -\int^h U(h')dh'$ and
\be
2{\cal F} - a\frac{\partial F}{\partial a} =
{\rm const}\cdot \int^h dh'' \exp\left(-\int^{h''}U(h')dh'\right)
+ {\rm const}'.
\label{Pic4}
\ee
In the particular case of (\ref{Pic2}) $\ U(h) = 0$, and the r.h.s.
of (\ref{Pic4}) is just $const\cdot h\ +\ const'$. The constants are
not fixed by Picard-Fuchs equations, since they depend on normalizations
of $d{\cal S}$ and the quasiclassical $\tau$-function $e^{{\cal F}}$.

 From (\ref{Pic2}) one can immediately obtain the asymptotics
of periods $\oint d{\cal S}_{\rm min}$. For large $u$ one can
neglect $\Lambda^4$ and get $\oint d{\cal S}_{\rm min} \sim u^\kappa$,
where $\kappa(\kappa -1) = -\frac{1}{4}$. This equation has
double root $\kappa = \frac{1}{2}$, therefore allowed asymptotics
are $\sqrt{u}$ and $\sqrt{u}\log u$ - in accordance with
(\ref{largeuaD}). For small $u$,
$$
\oint d{\cal S}_{\rm min} =
\sum_{n=0}^\infty \gamma_n\left(\frac{u}{\Lambda^2}\right)^n,
$$
and (\ref{Pic2}) provides the recurrent relation:
$$
\gamma_n = \frac{(n-\frac{1}{2})^2}{n(n-1)}\gamma_{n-2},
\ \ \ {\rm i.e.} \ \ \
\gamma_{2n} = \frac{4^n}{(2n)!}\left(\frac{\Gamma(n+\frac{1}{4})}
{\Gamma(\frac{1}{4})}\right)^2\gamma_0, \ \ \
\gamma_{2n+1} = \frac{4^n}{(2n+1)!}\left(\frac{\Gamma(n+\frac{3}{4})}
{\Gamma(\frac{3}{4})}\right)^2\gamma_1
$$

Though eqs. of the type (\ref{Pic1}) usually exist on moduli spaces of
compact complex surfaces,
their explicit form - as well as the possibility to lift them to some
identity for the prepotential ${\cal F}$ - is obscure in generic setting.
Therefore, generalization of above reasoning for higher-rank groups -
though possible - is rather involved, in any case it is far more
sophisticated than the approach based on the theory of prepotential.
Picard-Fuchs equations remain, however, the main available tool
when the spectral
surfaces are of non-unit complex dimension (like in the case with
Calabi-Yau manifolds) - and alternatives are still
underdeveloped.

\subsection{Toda-chain for $SL(N_c)$ \label{Todachain}}

We now proceed to example of pure gauge $N=2$  SUSY theory with the
gauge group $SU(N_c)$.
According to \cite{KLTY,AF,HO,MW1,IM}
the curve ${\cal C}$ in this case is of genus
$p = {\rm rank}(SL(N_c)) = N_c - 1$, and is given by the equation:
\be
z + \frac{1}{z} = 2P_{N_c}(t) \equiv
\Lambda^{-N_c} \sum_{l=0}^{N_c} s_l(h)t^{N_c - l}.
\label{Todacurve}
\ee
Here $s_l(h)$ are Schur polinomials, see \cite{IM} for details.
 It is important
that $s_1(h) = h_1 = 0$ for $SL(N_c)$ (but not for $GL(N_c)$).
Introducing $Y = \frac{1}{2}\left(z - \frac{1}{z}\right)$, one can
 rewrite this
in hyperelliptic form:
\be
Y^2 = P_{N_c}^2(t) -1
\ee
Holomorphic 1-differentials on the hyperelliptic curve are \cite{HE}
(linear combinations of)
 $dv_j = \frac{t^{j-1}}{Y(t)}dt$, $i \leq j \leq N_c-1 = p$.
(The basis $\{dv_j\}$ is not canonical, $\oint_{A_i}dv_j \neq \delta_{ij}$).
According to (\ref{diff}),
\be
d{\cal S}_{\rm min} \cong \frac{t}{i\pi }d\log z =
\frac{t}{i\pi }d\log\left(Y(t) + P_{N_c}(t)\right).
\label{Todadiff}
\ee
This indeed satisfies (\ref{dSdef}):
\be
i\pi \frac{\partial d{\cal S}_{\rm min}}{\partial h_k} \cong
td\frac{\partial P_{N_c}(t)/\partial h_k}{Y(t)} \cong
-\frac{\partial P_{N_c}(t)}{\partial h_k}\frac{dt}{Y(t)} = \nn \\
= -\frac{1}{2}\Lambda^{-N_c} \sum_{l=2}^{N_c}
\frac{\partial s_l(h)}{\partial h_k}\frac{t^{N_c - l}dt}{Y(t)} =
-\frac{1}{2}\Lambda^{-N_c}\sum_{j=1}^{N_c-1}
\frac{\partial s_{N_c+1-j}}{\partial h_k}dv_j(t).
\label{derShex}
\ee
The items with $l = 0,1$ (i.e. with $j = N_c, N_c+1$) are absent
 at the r.h.s.,
since $s_0(h) = 1$ and $s_1(h) = h_1 = 0$.\footnote{
Using the definition of Schur polinomials,
$\sum_l s_l(h)t^{-l} = \exp\left( -\sum_k h_kt^{-k}\right)$, and its
immediate corollary, $\frac{\partial s_l(h)}{\partial h_k} = -s_{l-k}(h)$,
the chain of identites (\ref{derShex}) can be further continued to give
$$
i\pi \frac{\partial d{\cal S}_{\rm min}}{\partial h_k} \cong
+\frac{1}{2}\Lambda^{-N_c} \sum_l s_{l-k}(h)\frac{t^{N_c-l}dt}{Y(t)}
= \left[t^{-k}P_{N_c}(t)\right]_+ \frac{dt}{Y(t)},
$$
where the projector is defined by
$\left[ \sum_{n=-\infty}^{+\infty} u_nt^n\right]_+ \equiv
\sum_{n=0}^{+\infty} u_nt^n$.
}

$d{\cal S}_{\rm min}$ has singularities at $z=0$ and $z=\infty$,
in both cases $t = \infty$, which is {\it not} a branching point.
The proper coordinate in the vicinity of singularities is $\xi = t^{-1}$,
and
$$
d{\cal S}_{\rm red}  \cong \frac{t}{i\pi}\frac{dz}{z} =
\frac{t}{i\pi}(d\log t^{N_c} + O(t^{-2})) =\frac{N_c}{i\pi}dt (1+ O(t^{-2})
= \frac{iN_c}{\pi}\frac{d\xi}{\xi^2}(1 + O(\xi^2)).
$$
 From this we conclude that $d{\cal S}_{\rm min}$ corresponds to the
choice $T_1 \sim {N_c}$ and all other $T_n = 0$.

In the particular case of $SL(2)$, the curve ${\cal C}_{SL(2)}$
considered in this section, does not coincide with that from
s.\ref{oriSW} (just the double ratios are different). Instead \cite{SW2} one
is the double cover of another, with the standard
KP$\leftrightarrow$Toda transformation
$\tilde x = -\frac{\Lambda^2}{2}\left(z + \frac{1}{z}\right)$.
The periods $a(h)$ and $a^D(h)$ are of course the same.

When $N_f$ matter $N=2$ supermultiplets in the fundamental
 representation of the gauge group are introduced, the equations
(\ref{Todacurve}) and (\ref{Todadiff}) change for \cite{HO}:
\be
z + \frac{1}{z} = \frac{2P_{N_c}(t)}{\sqrt{Q_{N_f}(t)}}
\ee
\be
d{\cal S}_{\rm min} \cong \frac{t}{i\pi}d\log z.
\ee
with the same $P_{N_c}(t)$ and a new polinomial $Q_{N_f}(t)$
of degree $N_f$, which depends on the masses of matter
multiplets.
This can be also rewritten as
\be
\hat z + \frac{Q_{N_f}(t)}{\hat z} = 2P_{N_c}(t),
\ee
but then
\be
d{\cal S}_{\rm min} =
\frac{t}{i\pi}d\log\left(\frac{\hat z}{\sqrt{Q_{N_f}(t)}}\right).
\ee
$d{\cal S}_{\rm min}$ now has extra simple poles, with
residues proportional to the masses \cite{SW2}. In such
case the integral (\ref{Freddef}) for the prepotential diverges
logarithmically, but the coefficients of divergent part are
pure quadratic in masses, thus integrals can be easily
regularized in controllable way. The periods are finite.

\sect{UV-finite $4d$ models and the $1d$ Calogero system:
Elliptic curve as the {\it bare} spectral surface  \label{Calod}}

The next set of examples is provided by gauge $N=2$ SUSY theories
with additional matter multiplet in the adjoint representation of the
gauge group \cite{SW2,DW}. As the mass $m$ of the mupltiplet
changes from $m = \infty$ to $m = 0$ the supersymmetry
increases from $N=2$ to $N=4$. In the framework of integrability
theory this flow is described by Calogero system \cite{M,GoM,IM}
with the role of $m$ played by the coupling constant $g \sim m$.
The curve ${\cal C}$ is represented as a  covering over
elliptic ''bare spectral curve'' $E(\tau)$.
This latter one can be parametrized
either with a ''flat'' coordinate $\xi$ (so that the relevant objects are
double periodic functions of $\xi$), or - alternatively - by elliptic
equation
\be
E(\tau): \ \ \ \ \ y^2 = \prod_{a=1}^3(x-\hat e_a(\tau)),
\label{Etaudef}
\ee
and
\be
d\omega_0 = \frac{d\xi}{w_1} = \frac{1}{2\pi}\frac{dx}{y(x)}.
\label{domega0}
\ee
Generic theory is rather sophisticated \cite{DW,IM}, and in this paper
we concentrate on the simplest case of the gauge group $SU(2)$.

\subsection{The case of the  $SL(2)$ group. Curve,
$d{\cal S}_{\rm min}$ and the prepotential}

In this case (\ref{curve}) says just:
\be
t^2 - h_2 = g^2\wp(\xi) = -\frac{m^2}{8}x,
\label{curcal}
\ee
thus ${\cal C}$ is a 2-sheet covering of $E(\tau)$, and
\be
d{\cal S}_{\rm min} \cong 2td\omega_0 =
\frac{2}{w_1}\sqrt{h_2 + g^2\wp(\xi)}d\xi =
\frac{1}{\pi}\frac{\sqrt{h_2 - \frac{m^2}{8}x}}{y(x)}dx.
\label{diffCalSl2}
\ee
Here
\be
\left(\frac{\pi}{w_1}\right)^2 x \equiv \wp(\xi) \equiv
\frac{1}{\xi^2} +
\sum_{\stackrel{-\infty < N,M <\infty}{N^2+M^2\neq 0}}
\left(\frac{1}{(\xi + Nw_1 + Mw_2)^2} -
\frac{1}{(Nw_1+Mw_2)^2}\right),
\label{wp}
\ee
is the doubly periodic Weierstrass $\wp$-function
on the ''bare spectral curve'' $E(\tau)$,
$\ \tau = \frac{w_2}{w_1}$, which satisfies the equation
\be
\frac{1}{4}\left[\wp'(\xi)\right]^2 =
\prod_{a=1}^3 (\wp(\xi) - e_a),
\ \ \ \ \hat e_a(\tau) \equiv
\left(\frac{w_1}{\pi}\right)^2 e_a(w_1,w_2),
\ \ \ \ \sum_{a=1}^3 \hat e_a(\tau) = 0
\label{wp'}
\ee
Note that our $w_{1,2}$ are periods, not half-periods of
$\wp(\xi)$. The coupling constant
$g$ is proportional to the mass $m$:
\be
g^2 = \frac{m^2}{8}\left(\frac{iw_1}{\pi}\right)^2.
\label{gverm}
\ee
See \cite{IM} for more details.

The entire curve ${\cal C}$
is a double covering of $E(\tau)$,
but periods of $d{\cal S}_{\rm min}$ - as is obvious from
(\ref{diffCalSl2}) - can be evaluated as if it was defined
on an auxiliary  genus-one curve
\be
\hat {\cal C}: \ \ \ \ \
\hat y^2 = (h_2 - \frac{m^2}{8} x)
\prod_{a=1}^3(x - \hat e_a(\tau)).
\label{hatC}
\ee
Obviously,
\be
d{\cal S}_{\rm min} =
\frac{1}{\pi}\frac{h_2-\frac{m^2}{8}x}{\hat y(x)}dx,
\label{CaldSmin}
\ee
and  $\frac{\partial d{\cal S}_{\rm min}}{\partial h_2} =
\frac{1}{2\pi}\frac{dx}{\hat y(x)}$
is a holomorphic 1-differential
on $\hat{\cal C}$ - in accordance with (\ref{dSdef}).
According to the
general principles of \cite{SW2}, if the mass $m$ of the
adjoint matter multiplet is non-vanishing, $\frac{m}{2\sqrt{2}}$
is the residue of
(\ref{diffCalSl2}) at its singularities (which are two simple poles,
both located at
$x = \infty$, but on different sheets of elliptic curve
$\hat {\cal C}$).

The (reduced) prepotential - as given by eq.(\ref{Freddef}) - is
\be
{\cal F}_{\tau}(a,m) = \frac{1}{2}aa_D +
\frac{im}{2\sqrt{2}}\int_{8h_2/m^2}^\infty \sqrt{
\frac{\frac{m^2}{8}x - h_2}{\prod_{a=1}^3(x - \hat e_a(\tau))}
}\ dx, \nn \\
a = \frac{2}{\pi}\int_{\hat e_1}^{\hat e_2} \sqrt{
\frac{h_2 - \frac{m^2}{8}x}{\prod_{a=1}^3(x - \hat e_a(\tau))}}\ dx,
\ \ \
a_D = \frac{2}{\pi}\int_{\hat e_2}^{\hat e_3} \sqrt{
\frac{h_2 - \frac{m^2}{8}x }{\prod_{a=1}^3(x - \hat e_a(\tau))}}\ dx.
\label{prepoCalSl2}
\ee
In this formula $\tau$ is a parameter, not an {\it argument}
of the prepotential. It is of interest also to consider it as
one of the arguments, on equal footing with the other moduli.
As usual in the presence of massive
matter multiplets the integral for the prepotential
is logarithmically divergent, the coefficient of the
logarithm is just $\frac{m^2}{8}$ and is independent of $h_2$
or $\tau$.

\subsection{The $m^2=0$ ($N=4$ SUSY) limit}

In this limit $g^2 \sim m^2 = 0$ and the curve ${\cal C}$
degenerates into (two copies) of the bare spectral curve:
\be
\hat{\cal C} \stackrel{m^2=0}{\longrightarrow} E(\tau)
\ee
while
\be
d{\cal S}_{\rm min} \cong 2td\omega_0
\stackrel{g^2=0}{\longrightarrow}
2\sqrt{h_2}d\omega_0,
\label{N=4diff}
\ee
- into the multiple of canonical holomorphic
$(1,0)$-differential on $E(\tau)$,
given by (\ref{domega0}).

The periods
\be
a \rightarrow  2\sqrt{h_2}, \ \ \ a_D \rightarrow 2\tau\sqrt{h_2},
\ee
and the prepotential  (\ref{prepoCalSl2}) turns into
\be
{\cal F}_\tau(a) = \frac{1}{2}aa_D = \frac{1}{2}a^2\tau
= 2h\tau.
\ee
All this is in accordance with  \cite{SW2}.

Description of the $N=4$ SUSY limit $g^2 \sim m^2 = 0$
is equally simple for any group $SL(N_c)$. Its
characteristic feature is that
the spectral parameter $z$
disappears from the Lax operator in eq.(\ref{curve}) for the
spectral curve ${\cal C}$ and (\ref{curve}) turns into:
\be
P_{N_c}(t) \equiv \sum_{l=0}^{N_c} s_l(h)t^{N_c - l} = 0,
\ee
solutions to which are just $t = t_\gamma$, $\ \gamma = 1,\ldots,N_c$,
such that $h_k = \frac{1}{k}\sum_{\gamma = 1}^{N_c}t_\gamma^k$
(and $h_1 = \sum_\gamma t_\gamma = 0$ for $SL(N_c)$).
We see that the curve ${\cal C}$ splits into $N_c$ copies (glued
together at the points $t_\gamma = t_{\gamma'}$, where the
Hamiltonians $h_k$ become algebraically dependent).
The differential
\be
d{\cal S}_{\rm min} \cong \oplus_\gamma t_\gamma d\omega_0(z),
\ee
it is free of singularities, thus all the corresponding $T_n = 0$.
Its periods are just
\be
a_\gamma = t_\gamma\oint_{A_0}d\omega_0 = t_\gamma, \ \ \
a^D_\gamma = t_\gamma\oint_{B_0}d\omega_0 = \tau t_\gamma
\ee
(one can take any $N_c-1$ of these for $a_i$ and $a_i^D$).
The prepotential in this case is
\be
{\cal F}_{\rm red}(a_i) =
\frac{1}{2}\sum_\gamma a_\gamma a^D_\gamma
= \frac{1}{2}\tau\sum_\gamma a_\gamma^2,
\ee
it is obviously homogeneous of degree 2, since $\tau$ is
a parameter of $E(\tau)$, which is by definition independent of
$a_i$ - and this is in accordance with both conformal invariance of
the $N=4$ SUSY theory and with eq.(\ref{homF}) when all $T_n = 0$.

\subsection{The double scaling $m^2=\infty$ (pure gauge $N=2$
or sine-Gordon) limit \label{scali}}

The limit $g^2 \sim m^2 \rightarrow \infty$ is more sophisticated.
Of interest is actually a {\it double scaling} limit \cite{Ino,IM},
when also $q = e^{i\pi\tau} \rightarrow 0$, so that dimensional
trasmutation takes place and the new massive parameter
emerges,
\be
\Lambda^2 = 2m^2q.
\label{Lambdadef}
\ee
The scaling rule for $\xi$ is:
\be
 \xi = \frac{iw_1}{2\pi}\log\frac{z}{q}, \ \ \
\frac{2}{w_1}d\xi = \frac{i}{\pi}\frac{dz}{z}.
\ee
We have:
\be
\wp(\xi) = \left(\frac{\pi}{iw_1}\right)^2
\left(C(\tau) + 4q\left(z + \frac{1}{z}\right) + o(q^2)\right),
\ee
where
\be
C(\tau) = \frac{1}{3}\frac{\partial \log\Delta(q^2)}
{\partial \log q^2} = \frac{1}{3}\left(1 - 24
\sum_{M\geq 1}\frac{q^{2M}}{(1 - q^{2M})^2}\right)
= \frac{1}{3}\left(1 - 24q^2 - 72q^4 -  96q^6 -\ldots\right),
\label{Cfdef}
\ee
\be
\Delta(q^2) \equiv \eta^{24}(q^2) \equiv q^2\prod_{n\geq 1}
(1 - q^{2n})^{24} = \frac{9}{256}(\hat e_1-\hat e_2)^2
(\hat e_2-\hat e_3)^2(\hat e_3-\hat e_1)^2.
\ee
Therefore
\be
h_2 + g^2\wp(\xi) =
\left( h_2 + \frac{m^2}{8}C(\tau)\right) \ + \
\frac{1}{2}m^2q\left(z + \frac{1}{z}\right) + o(q^2) \equiv
\nn \\
\equiv
\frac{1}{2}\left( u + \frac{1}{2}\Lambda^2 \left(z + \frac{1}{z}\right)
+ o(q^2)\right).
\ee
The double scaling limit is the one when $m \rightarrow \infty$,
$q\rightarrow 0$ (i.e. $\tau \rightarrow +i\infty$),
$h = h^{N=4} \rightarrow \infty$,
but  $\Lambda^2$ and\footnote{
It deserves noting that $C(\tau) \neq \frac{1}{2}\hat e_1(\tau)
= 2C(2\tau) - C(\tau)$, and thus  (\ref{h2h4})
is different from eq.(16.25) of \cite{SW2}. See \cite{IM}
for more details.
}
 \be
u \equiv 2h_2^{N=2} \equiv
2\left(h_2^{N=4} + \frac{m^2}{8}C(\tau)\right)
\label{h2h4}
\ee
remain finite. Then
$$
h_2 + g^2\wp(\xi) \ \stackrel{{\rm sc.l.}}{\longrightarrow}\
\frac{u}{2} + \frac{\Lambda^2}{4}\left( z + \frac{1}{z}\right)
= \frac{1}{2}(u - \tilde x),
$$
where  $\frac{2}{w_1}d\xi =
\tilde x \equiv -\frac{\Lambda^2}{2}\left( z + \frac{1}{z}\right)$,
so that  $\frac{i}{\pi} \frac{dz}{z} = \frac{i}{\pi}
\frac{d\tilde x}{\sqrt{\tilde x^2 - \Lambda^4}}$.

Given all this,
\be
d{\cal S}_{\rm min} = \frac{2}{w_1}\sqrt{h_2 + g^2\wp(\xi)}d\xi
\ \stackrel{{\rm sc.l.}}{\longrightarrow}\  \frac{1}{\pi\sqrt{2}}
\sqrt{\frac{\tilde x-u}{\tilde x^2 - \Lambda^4}}\ d\tilde x,
\ee
in accordance with (\ref{dSexa1}) and \cite{SW}.

\subsection{The double scaling limit  in elliptic parametrization}

It is instructive to consider the same limit using elliptic
parametrization. This helps to follow degeneration of
$\hat{\cal C}$ into a punctured sphere.
As we found in the previous section
\ref{scali}, the coordinate $x$ on $E(\tau)$ in the double scaling
limit is substituted by $\tilde x$:
\be
x = \left(\frac{w_1}{\pi}\right)^2\wp(\xi) =
- \left(C(\tau) + 4q\left(z + \frac{1}{z}\right) + o(q^2)\right)
= -\frac{1}{3} + \frac{8q}{\Lambda^2}\tilde x + o(q^2),
\ee
so that $dx = \frac{8q}{\Lambda^2}d\tilde x (1 + o(q))$.
Equation (\ref{Etaudef}) for $E(\tau)$ contains also
$\hat e_a(\tau)$, which are expressible in terms of
the theta-constants:
\be
\hat e_1(\tau) = \frac{2\theta_{00}^4 - \theta_{10}^4}{3} =
\frac{2}{3}( 1 + 24q^2 + 24q^4 + \ldots)
= \frac{2}{3} + o(q^2), \nn \\
\hat e_2(\tau) = -\frac{\theta_{00}^4 + \theta_{10}^4}{3} =
-\frac{1}{3}(1+ 24 q + 24q^2 + 96q^3 + 24q^4 + 144q^5 + \ldots)
= -\frac{1}{3} - 8q + o(q^2), \nn \\
\hat e_3(\tau) = \frac{2\theta_{10}^4 - \theta_{00}^4}{3} =
-\frac{1}{3}(1- 24 q + 24q^2 - 96q^3 + 24q^4 - 144q^5 + \ldots)
= -\frac{1}{3} + 8q + o(q^2)
\ee
Thus $E(\tau)$ becomes:
\be
y^2 = \prod_{a=1}^3(x - \hat e_a(\tau)) =
-\frac{64q^2}{\Lambda^4}(\tilde x^2 - \Lambda^4)(1 + o(q))
\ee
and
\be
\frac{dx}{y(x)} = -i\frac{d\tilde x}{\sqrt{\tilde x^2 - \Lambda^4}}
(1 + o(q)).
\ee
When multiplied by $\frac{1}{\pi}\sqrt{h - \frac{m^2}{8}x}
= \frac{i}{\sqrt{2}\pi}\sqrt{\tilde x - u}\ (1 + o(q))$, this once again
reproduces (\ref{dSexa1}).

\subsection{Picard-Fuchs equations for $\oint d{\cal S}_{\rm min}$}

Let us denote
$$
<(\ldots)> \ \equiv \oint dx\sqrt{\frac{\hat h -x}{\prod_{a=1}^3
(x - \hat e_a(\tau))}}\ (\ldots) = \oint d\hat{\cal S}_{\rm min}(\ldots).
$$
Then $a $ and $a^D$ are given by
$\frac{m}{2\sqrt{2}\pi}< 1 >$, and $\hat h \equiv \frac{8h_2}{m^2}$,
$d\hat{\cal S}_{\rm min} \equiv
\frac{2\sqrt{2}\pi}{m}d{\cal S}_{\rm min}$.
The Ward identity:
$$
<\left(\frac{1}{\hat h-x} + \sum_{a=1}^3\frac{1}{x-\hat e_a}\right)
\delta x >\ = 2<\frac{d\delta x}{dx}>
$$
holds for any $\delta x$.
Its particular examples are:
\be
\begin{array}{ll}
\delta x = 1: &
 <\left(\frac{1}{\hat h-x} +
\sum_{a=1}^3\frac{1}{x-\hat e_a}\right)>\ = 0, \\
\delta x = x:  &
<\left(\frac{\hat h}{\hat h-x} +
\sum_{a=1}^3\frac{\hat e_a}{x-\hat e_a}\right)>
\ = 0, \\
\delta x = \frac{1}{\hat h - x}:  &
<\frac{1}{(\hat h-x)^2}>\ =
\sum_{a=1}^3<\frac{1}{(x-\hat e_a)(\hat h-x)}> \ = \\
& \ \ = \left(\sum_{a=1}^3\frac{1}{\hat h - \hat e_a}\right)
<\frac{1}{\hat h-x}> +
\sum_{a=1}^3\left(
\frac{1}{\hat h-\hat e_a}<\frac{1}{x-\hat e_a}>\right)
\end{array}
\ee
Combining the first and the third of these relations we get:
\be
<\frac{1}{(\hat h-x)^2}>\  - \frac{2}{3}
\left(\sum_{b=1}^3 \frac{1}{\hat h - \hat e_b}\right)
<\frac{1}{ \hat h - x} > =
\sum_{a=1}^3
<\frac{E_a}{x-\hat e_a}>, \nn \\
E_a(\hat h|\tau)  \equiv
\frac{1}{\hat h-\hat e_a} - \frac{1}{3}\left(\sum_{b=1}^3
\frac{1}{\hat h-\hat e_b}\right), \ \ \ \
\sum_{a=1}^3 E_a = 0.
\ee
This list should be supplemented by ($q = e^{i\pi\tau}$):
\be
\frac{\partial}{\partial \log q}<\ 1\ >\ =
\frac{1}{2}\sum_{a=1}^3
<\frac{\partial \hat e_a/\partial\log q}{x-\hat e_a}>
\ee
Relation (it can be considered as a definition of $f(\hat h|\tau)$
and $g(\hat h|\tau)$, see also (\ref{edot}) below)
\be
\frac{\partial \hat e_a(\tau)}{\partial \log q} =
 f(\hat h|\tau)\hat e_a(\tau) +
g(\hat h|\tau)E_a(\hat h|\tau)
\label{reledot}
\ee
implies the Picard Fuchs equation for $d{\cal S}_{\rm min}$
in the form:
\be
\left(\frac{\partial}{\partial \log q} + \hat hf(\hat h|\tau)
\frac{\partial}{\partial\hat h} + 2g(\hat h|\tau)
\left[\frac{\partial^2}{\partial\hat h^2} + \frac{1}{3}
\left(\sum_{b=1}^3 \frac{1}{\hat h - \hat e_b}\right)
\frac{\partial}{\partial\hat h}\right]\right) d\hat{\cal S}_{\rm min}
\cong 0.
\label{PFCal1}
\ee

Now we need to find $f(\hat h|\tau)$ and $g(\hat h|\tau)$
from (\ref{reledot}). Of these three equations only two are
independent, since the sum of the three is zero.
We begin from rewriting $E_a$ in a more convenient
form:
$$
E_a = \frac{3\hat h\hat e_a + \hat e_a^2 +
2\varepsilon_{abc}\hat e_b\hat e_c}
{3y^2(\hat h)} = \frac{\hat h\hat  e_a + \hat e_a^2 -
\frac{1}{3}\sum_{b=1}^3 \hat e_b^2}{y^2(\hat h)},
\ \ \  \
\hat e_aE_b - \hat e_bE_a =
- \frac{(\hat e_a - \hat e_b)(\hat e_a\hat e_b+
2\varepsilon_{abc}\hat e_c^2)}{3y^2(\hat h)}.
$$
Here $y^2(\hat h) = {\prod_{a=1}^3(\hat h - \hat e_a(\tau))}$,
and $\varepsilon_{123} =
\varepsilon_{213} = \ldots =1$, $\varepsilon_{112} = \ldots = 0$,
with {\it no} summation over repeated indices.
Using these formulas we obtain from (\ref{reledot}):
\be
g(\hat h|\tau) = \frac{-\frac{\partial\hat e_1}{\partial \log q}\hat e_2
+ \frac{\partial\hat e_2}{\partial \log q}\hat e_1}
{\hat e_1E_2 - \hat e_2E_1} = 3y^2(\hat h)\hat g =
- y^2(\hat h), \nn \\
f(\hat h|\tau) = \frac{\frac{\partial\hat e_1}{\partial \log q}E_2
- \frac{\partial\hat e_2}{\partial \log q}E_1}
{\hat e_1E_2 - \hat e_2E_1} = -(3\hat h\hat g + \hat f) =
\hat h + C(\tau).
\label{PicCalaux}
\ee
In more detail, the coefficients at the r.h.s. are:
\be
\hat g \equiv  \frac{\frac{\partial\hat e_1}{\partial \log q}\hat e_2
- \frac{\partial\hat e_2}{\partial \log q}\hat e_1}
{(\hat e_1-\hat e_2)(\hat e_1\hat e_2 + 2\hat e_3^2)} =
\frac{\frac{16}{3}q( 1 - 12 q^2 + 54 q^4 - \ldots)}
{-16q(1 - 12q^2 + 54q^4 - \ldots)} = -\frac{1}{3}, \nn \\
\hat  f  \equiv \frac{\frac{\partial\hat e_1}{\partial \log q}
(2\hat e_1\hat e_3 + \hat e_2^2) -
 \frac{\partial\hat e_2}
{\partial \log q}(2\hat e_2\hat e_3 + \hat e_1^2)}
{(\hat e_1-\hat e_2)(\hat e_1\hat e_2 + 2\hat e_3^2)} =
\frac{\frac{16}{3}q( 1 - 36 q^2  + 270 q^4 - \ldots)}
{-16q(1 - 12q^2 + 54q^4 - \ldots)} = \nn \\
= -\frac{1}{3}(1 - 24q^2 - 72q^4 - \ldots) = -C(\tau) =
-\frac{1}{3}\frac{\partial\log \Delta(q^2)}{\partial \log q^2}.
\ee
Note that $C(\tau)$ here is exactly the same as in (\ref{Cfdef}).

As additional check, one can rewrite (\ref{reledot}) as
\be
\frac{\partial\log\theta_{00}^4}{\partial\log q} =
\frac{\partial\log(\hat e_1 - \hat e_2)}{\partial\log q} =
f(\hat h|\tau) + g(\hat h|\tau)\frac{h-\hat e_3}{y^2(\hat h)}
\label{inter1}
\ee
plus two analogous
relations for $\theta_{10}$ and $\theta_{01}$.
Adding all the three one gets,
\be
f(\hat h|\tau) + \frac{hg(\hat h|\tau)}{y^2(\hat h)}
= \frac{1}{3}\frac{\partial\log
(\theta_{00}^4\theta_{10}^4\theta_{01}^4)}{\partial\log q} =
\frac{1}{3}\frac{\partial \log\Delta(q^2)}{\partial \log q^2} =
C(\tau),
\ee
(since $\Delta(q^2) = (\theta_{00}\theta_{10}\theta_{01})^8 =
(\theta_{11}')^8$).
Another linear combination of formulas (\ref{inter1}) gives:
\be
g(\hat h|\tau) = \frac{y^2(\hat h)}{\theta_{01}^4}
\frac{\partial\log(\theta_{00}^4/\theta_{10}^4)}{\partial\log q}
= -y^2(\hat h).
\ee
Of course, when (\ref{PicCalaux}) are substituted into
(\ref{reledot}), the $\hat h$-dependence cancels at the r.h.s.:
\be
\frac{\partial \hat e_a}{\partial \log q} = \frac{{\rm g}_2}{6}
+ C\hat e_a - \hat e_a^2,
\label{edot}
\ee
where
\be
{\rm g_2}(\tau) = -4(\hat e_1\hat e_2 + \hat e_2\hat  e_3
+ \hat e_3\hat e_1) = 2\sum_{a=1}^3 \hat e_a^2 =
\frac{2}{3}(\theta_{00}^8 + \theta_{01}^8 + \theta_{10}^8) = \nn \\
= \frac{4}{3}\left(1 + 240(q^2 + 9q^4 + 28 q^6 + \ldots)\right) =
\frac{4}{3}\left(1 + 240q^2 + 2160 q^4 + 6720 q^6 + \ldots\right).
\ee

Substituting (\ref{PicCalaux}) into (\ref{PFCal1}),  we obtain
\be
\frac{1}{2}\frac{\partial}{\partial\log q} d\hat{\cal S}_{\rm min}
\cong \left(y^2(\hat h)\frac{\partial^2}{\partial \hat h^2} +
\left[\frac{1}{2}\hat h^2 - \frac{1}{2}\hat hC(\tau) -
\frac{1}{12}{\rm g}_2(\tau)\right]\frac{\partial}
{\partial\hat h}\right)d\hat{\cal S}_{\rm min}
\label{PicCal}
\ee


\subsection{The $m^2 \rightarrow \infty$, $q \rightarrow 0$
double scaling ($N=2$) limit of Picard-Fuchs equation}

We first consider the  double scaling limit, $m^2 \rightarrow
\infty$, $q \rightarrow 0$. For small $q$ the operator at the r.h.s.
of (\ref{PicCal}) becomes:
\be
\left(\left(\hat h + \frac{1}{3}\right)^2 - 8q^2\right)
\left(\hat h - \frac{2}{3}
\right)\frac{\partial^2}{\partial \hat h^2} \ +
\ \frac{1}{2}\left(\hat h + \frac{1}{3}\right)
\left(\hat h - \frac{2}{3}\right)
\frac{\partial}{\partial \hat h}
\label{rhsPFECaldsl}
\ee
(one can use the above-listed asymptotic expansions for
$\hat e_a(\tau)$ and $C(\tau)$).
According to (\ref{h2h4}) and (\ref{Lambdadef})
these are $\hat h \equiv \frac{8h}{m^2}$, $u$ and $\Lambda$,
that are finite in the double scaling limit, and
\be
\hat h + \frac{1}{3} = \hat h + C(\tau) + o(q^2) =
\frac{4u}{m^2} + o(q^2) = \frac{8u}{\Lambda^2}q + o(q^2).
\ee
After this substitution (\ref{rhsPFECaldsl}) becomes:
\be
(\Lambda^4 - u^2)\frac{\partial ^2}{\partial u^2} -
\frac{1}{2}u\frac{\partial}{\partial u}.
\label{lim23}
\ee
Now, turn to the l.h.s. of (\ref{PicCal}). Obviously,
\be
\left.\frac{1}{2}\frac{\partial}{\partial\log q}\right|_{h = {\rm const}}
= \frac{1}{2}\left(
\left.\frac{\partial}{\partial\log q}\right|_{u = {\rm const}} -
\left[\hat h + \frac{1}{3}\right]\frac{\partial}{\partial\hat h}\right) =
\left.\frac{1}{2}\frac{\partial}{\partial\log q}\right|_{u = {\rm const}}
- \frac{1}{2}u\frac{\partial}{\partial u}.
\ee
The second item at the r.h.s. is exactly what necessary to cancel the
one at the
r.h.s. of (\ref{lim23}). The $q$-derivative should be irrelevant in
the limit, when it acts on $d{\cal S}_{\rm min}$, but (\ref{PicCal})
is written for $d\hat{\cal S}_{\rm min} \sim
\frac{1}{m}d{\cal S}_{\rm min} \sim q^{1/2} d{\cal S}_{\rm min}$.
Thus we should substitute $\frac{1}{2}\frac{\partial}{\partial\log q}
d\hat{\cal S}_{\rm min} \rightarrow \frac{1}{4}d\hat{\cal S}_{\rm min}$.
Putting all together we obtain as the double-scaling limit of
eq.(\ref{PicCal}):
\be
\left((u^2 - \Lambda^4)\frac{\partial^2}{\partial u^2} + \frac{1}{4}
\right)d{\cal S}_{\rm min} \cong 0,
\ee
exactly the right Picard-Fuchs equation (\ref{Pic2}).

\subsection{The $m^2 = 0$ ($N=4$) limit}

Proceed now to another  limit, when  $m^2 = 0$.
In the first approximation, we can just say that
$\hat h \equiv \frac{8h}{m^2}$ is larger than everything in
this limit and (\ref{PicCal}) reduces to:
\be
\left(\hat h^3\frac{\partial^2}{\partial\hat h^2} +
\frac{1}{2}\hat h^2\frac{\partial}{\partial\hat h}\right)
d{\cal S}_{\rm min} \cong 0,
\ee
what gives the answer $d{\cal S}_{\rm min} \sim \sqrt{\hat h}$ -
in consistency with (\ref{N=4diff}), $d{\cal S}_{\rm min} \cong
2\sqrt{h}d\omega_0$. This accuracy, however, is not enough to
determine the $\tau$ dependence of the periods ($1$ and $\tau$)
of $d\omega_0$. Moreover, one can wonder how two such
different periods can satisfy the same equation.
Resolution to this puzzle is simple:
to get  the next approximation we
restore $m^2$-dependence in (\ref{PicCal}) and substitute
$d{\cal S}_{\rm min} = 2\sqrt{h_2}dS_0(\tau) -
\frac{m^2}{8\sqrt{h_2}}dS_1(\tau)$:
\be
\left(\frac{\partial}{\partial\log q} + \frac{1}{2}C(\tau)\right)
dS_0(\tau) \cong
-\frac{1}{2}dS_1(\tau),
\label{inter21}
\ee
and this is not a closed equation for $dS_0(\tau)$ alone.

One can check that $dS_0 = d\omega_0$ and
$dS_1 = xd\omega_0$ - as implied by $\frac{1}{h_2}$-expansion
of $d{\cal S}_{\rm min}$ (\ref{CaldSmin}) - do indeed
satisfy (\ref{inter21}). This follows from the Ward
identities for the average
$$
\langle\langle\ (\ldots)\ \rangle\rangle
 \equiv \oint d\omega_0\ (\ldots) =
\frac{1}{2\pi}\oint\frac{dx}{\sqrt{\prod_{a=1}^3(x-\hat e_a)}}
\ (\ldots),
$$
\be
\label{WIsm}
\begin{array}{ll}
\delta x = 1: & \sum_{a=1}^3 \langle\langle\frac{1}{x-\hat e_a}
\rangle\rangle \ = 0, \\
\delta x = x: &  \sum_{a=1}^3 \langle\langle\frac{\hat e_a}{x-\hat e_a}
\rangle\rangle \ = -\langle\langle 1 \rangle\rangle, \\
\delta x = x^2: & \sum_{a=1}^3 \langle\langle\frac{\hat e_a^2}{x-\hat e_a}
\rangle\rangle \ = \langle\langle x \rangle\rangle,
\end{array}
\ee
together with (\ref{edot}):

\be
\frac{\partial}{\partial\log q}\oint  d\omega_0 = \frac{1}{2}
\sum_{a=1}^3 \langle\langle\frac{\partial \hat e_a/\partial\log q}
{x-\hat e_a}\rangle\rangle \  \stackrel{(\ref{edot})}{=} \nn \\
= \frac{{\rm g}_2}{12}\sum_{a=1}^3 \langle\langle\frac{1}{x-\hat e_a}
\rangle\rangle \  + \frac{C(\tau)}{2}
\sum_{a=1}^3 \langle\langle\frac{\hat e_a}{x-\hat e_a}\rangle\rangle \
- \frac{1}{2}\sum_{a=1}^3 \langle\langle\frac{\hat e_a^2}{x-\hat e_a}
\rangle\rangle \ \stackrel{(\ref{WIsm})}{=} \nn \\
= -\frac{1}{2}\langle\langle x \rangle\rangle - \frac{C(\tau)}{2}
\langle\langle 1 \rangle\rangle\ = -\frac{1}{2}\oint xd\omega_0
- \frac{C(\tau)}{2}\oint d\omega_0,
\ee
i.e. indeed, as required by (\ref{inter21}), for any contour
\be
\left(\frac{\partial}{\partial\log q} + \frac{1}{2}C(\tau)\right)
\oint d\omega_0 = -\frac{1}{2}\oint xd\omega_0.
\ee
This identity can be also reinterpreted as  the vanishing of
residues of Weierstrass function, so that $\zeta$-function,
$\zeta(\xi) = -\int^\xi\wp(\xi)d\xi$, locally exists, and
\be
\oint_A xd\omega_0 = \frac{w_1}{\pi^2}\int_0^{w_1}
\wp(\xi)d\xi = \frac{w_1}{\pi^2}\left(\zeta(\xi) - \zeta(\xi + 2w_1)\right)
\equiv -\frac{2w_1\eta_1}{\pi^2} = \nn \\
= -\left(C(\tau) + 2\frac{\partial}
{\partial \log q}\right)\oint_Ad\omega_0 = -C(\tau), \ \ \
\oint_B xd\omega_0 = -\frac{2w_1\eta_2}{\pi^2} =
-\frac{2}{i\pi} - \tau C(\tau).
\ee

It deserves noting, that eq.(\ref{inter21}) and thus (\ref{PicCal})
has many other solutions, for example:
\be
d{\cal S}_{\rm min}\left(1 + o(h_2^{-2})\right)
dS_0 \sim \Delta^{-1/12}(q^2)\sqrt{h_2} =
\frac{\sqrt{h_2}}{{Det}\ \bar\partial_0}, \ \ \ dS_1 = 0.
\ee
Such extra solutions are eliminated by the second Picard-Fuchs
equation, which is not considered in this paper.

\subsection{Canonical form of the Picard-Fuchs equation
for the case of elliptic {\it bare} spectral curve \label{CafPF}}

The form of the r.h.s. of (\ref{PicCal}) implies that the ''proper''
variable in it is not $\hat h$, but rather $\chi$, such that
$d\chi \sim d\hat h/y(\hat h)$, or $\hat h = \wp(\chi)$.
More precisely, one can perform a conjugation,
$d{\cal S}_{\rm min} = U(\hat h,\tau)d\check{\cal S}$
and convert (\ref{PicCal}) into:
\be
\frac{1}{2}\frac{{\cal D}}{{\cal D}\log q}d\check{\cal S} \cong
\left[\left(y(\hat h)\frac{\partial}{\partial\hat h}\right)^2 +
V(\hat h,\tau)\right] d\check{\cal S}.
\label{PicCalcan}
\ee
We are not aware of exact form of the conjugation trasform
and thus of $V(\hat h,\tau)$ in (\ref{PicCalcan}), thus we
present only some preliminary formulas.

The most reasonable candidate for the role of $U(\hat h,\tau)$
seems to be
\be
U(\hat h,\tau) = \left(\hat h - 2C(\tau)\right)^{1/8}y^{1/4}(\hat h) =
\left[(\hat h - 2C(\tau))\prod_{a=1}^3(\hat h
- \hat e_a(\tau))\right]^{1/8}.
\ee
After such conjugation the $\partial/\partial \hat h$-term in
(\ref{PicCal}) is not completely eliminated - in order to
give the r.h.s. of (\ref{PicCalcan}),- but turns into
\be
\left(\frac{\hat h^2}{2} - \frac{C\hat h}{2} - \frac{{\rm g}_2}{12}
+ 2y^2(\hat h)\frac{\partial\log U}{\partial\hat h} -
\frac{\partial\log y(\hat h)}{\partial\hat h}\right)\frac{\partial}
{\partial\hat h} = \nn \\
= \left(\frac{1}{4}y^2(\hat h)\left(\frac{1}{\hat h - 2C} +
\sum_{a=1}^3\frac{1}{\hat h - \hat e_a}\right) -
\left(\hat h^2 + \frac{C\hat h}{2} - \frac{{\rm g}_2}{24}\right)
\right)\frac{\partial}{\partial \hat h} = \nn \\
= \frac{1}{4}\left(\frac{\hat h^3 - \frac{1}{4}{\rm g}_2\hat h
- \frac{1}{4}{\rm g}_3}{\hat h - 2C} -
\left(\hat h^2 + 2C\hat h + \frac{{\rm g}_2}{12}\right)\right)
\frac{\partial}{\partial\hat h}
\label{inter31}
\ee

What is special about this sophisticated expression  is that its
$\tau$-{\it in}dependent part vanishes. To be precise,
near $q = 0$ the r.h.s. of (\ref{inter31}) becomes
\be
\left(\frac{\left(\hat h + \frac{1}{3}\right)^2\left(\hat h -
\frac{2}{3}\right)}{\hat h - \frac{2}{3}} - \left(\hat h^2 +
\frac{2}{3}\hat h + \frac{1}{9}\right) + o(q^2)\right)\frac{\partial}
{\partial\hat h} = o(q^2)\frac{\partial}{\partial\hat h}.
\ee
This, in turn, implies that such term {\it can} be absorbed into
the l.h.s. of (\ref{PicCal}) by the following trick. In (\ref{PicCal})
the $(\log q)$-derivative is taken at constant $\hat h$.
Imagine now that we instead keep constant some other
variable, $H = H(\hat h,\tau)$ (ideally $H \sim \chi$, but this
is not guaranteed by our reasoning). Then
\be
\frac{\cal D}{{\cal D}\log q} \equiv \left.\frac{\partial}{\partial\log q}
\right|_{H = {\rm const}} = \left.\frac{\partial}{\partial\log q}
\right|_{\hat h} + \frac{\partial H}{\partial\log q}\frac{\partial}
{\partial H} = \frac{\partial}{\partial\log q} +
\frac{\partial H}{\partial\log q}\frac{\partial\hat h}{\partial H}
\frac{\partial}{\partial\hat h}.
\ee
Specific feature of such correction is that (as long as $H$
is expandable in powers of $q$) it vanishes at $q=0$ -
and this is exactly what (\ref{inter31}) does.

One can proceed a little further, considering
$\frac{1}{\hat h}$-expansion of (\ref{inter31}) - though we
restrict our consideration to the first non-trivial term of this
expansion. The r.h.s. of (\ref{inter31}) is actually:
\be
\left[\left(C^2(\tau) - \frac{{\rm g}_2(\tau)}{12}\right) +
o\left(\frac{1}{\hat h}\right)\right]\frac{\partial}{\partial\hat h}.
\ee
Of course, $C^2 - \frac{1}{12}{\rm g}_2 = o(q^2)$, as all the
other coefficients in the $\hat h^{-1}$-expansion. Moreover, it
is easy to check that
\be
C^2 - \frac{{\rm g}_2}{12} = 2\frac{\partial C}{\partial\log q},
\ee
so that
\be
\frac{\partial H}{\partial\log q}\frac{\partial\hat h}{\partial H} =
- 4\frac{\partial C}{\partial\log q} + o\left(\frac{1}{\hat h}\right)
\ee
and $ H \sim e^{-4C(\tau)}\hat h^{r}\left({\rm const} + o(\hat h^{-1})
\right)$
(the value of  $r$ does not affect our reasoning, if $H \sim
e^{-4C(\tau)}\chi$ we need $r = -\frac{1}{2}$).

Finally, to get some impression of what $V(\hat h)$ in
(\ref{PicCalcan}) can look like, evaluate it with our
$U(\hat h) = \left(\hat h - \frac{2}{3}\right)^{1/4}\left(\hat h +
\frac{1}{3}\right)^{1/4} + o(q^2)$ at $q=0$. The answer is:
\be
V(\hat h) = U^{-1}\left[\left(\hat h + \frac{1}{3}\right)^2
\left(\hat h - \frac{2}{3}\right)\frac{\partial^2}{\partial\hat h^2} +
\frac{1}{2}\left(\hat h + \frac{1}{3}\right)
\left(\hat h - \frac{2}{3}\right)\frac{\partial}{\partial\hat h} + o(q^2)
\right]U = \nn \\
= -\frac{1}{8}\frac{\hat h + \frac{5}{6}}{\hat h - \frac{2}{3}} + o(q^2).
\ee
This formula implies that one should actually perform a rational
transformation $\hat h \rightarrow \check h = -\frac{1}{8}
\frac{\hat h + \frac{5}{6}}{\hat h - \frac{2}{3}} + o(q^2)$;
$\ y(\hat h) \rightarrow \check y (\check h)$ before switching
to the new variables like $\chi$ or $H$ (rational transformation
preserve the shape of the vector field $y(\hat h)\partial/\partial\hat h$).

To summarize, hypothetically, in appropriate coordinates
(\ref{PicCal})  acquires ''canonical'' form:
\be
\frac{1}{2}\frac{{\cal D}}{{\cal D}\log q}
d\check{\cal S} \cong
\left(\left(\check y(\check h)\frac{\partial}{\partial \check h}\right)^2
+ \check h\right)d\check{\cal S} = \left(
\frac{\partial^2}{\partial \chi^2} - g^2\wp(\chi)\right)d\check{\cal S}.
\label{PicCaldream}
\ee

\subsection{Comments \label{CoPic}}

A few more comments are now in order about the Picard-Fuchs
eq.(\ref{PicCal}).

First of all, since we are now in the situation with two moduli,
the single equation is not enough to fix the cohomology
class of $d{\cal S}_{\rm min}$ unambiguosly: another one
is also required. It is not a big problem, but it should be kept
in mind in the future work on this subject.

Second, somewhat similar equations were recently introduced
in \cite{Ol}. This whole subject is of course intimately related
to the theory of Knizhnik-Zamolodchikov equation.

Third, the appearence of (\ref{PicCal}) is reminiscent of
the Schroedinger equation, especially
if it can indeed be brought to the form like (\ref{PicCaldream}):
then it is just the Schroedinger equation for $1d$ Calogero model.
Therefore, if successful, the derivation of (\ref{PicCaldream})
could serve as an illustration of the general
principle that the Witham method is essentially the same as
quantization - but with considerable change in the nature of
variables: quantized model lives on the {\it moduli} space
(the one of ''zero-modes'' or ''collective coordinates''),
not on original  configuration space.

Forth, if (\ref{PicCal}) can indeed be converted into
eq.(\ref{PicCaldream}), this would allow
one to interpret the modulus $\hat h$ as belonging
to the bare spectral surface $E(\tau)$ (while in the
original Seiberg-\-Witten/sine-Gordon setting it rather
belonged to a fundamental domain of $\Gamma_2$ on the
upper half-plane).
In other words, the full moduli space $\{\hat  h, \tau\}$ at every
given $\tau$ reduces to elliptic curve $E(\tau)$.

These remarks are already enough to explain why eq.(\ref{PicCal})
(and its analogues for other groups) should be investigated more
deeply. We are going now to add one more direction: it should
be compared to  Picard-Fuchs equation for the
relevant Calabi-Yau manifold - and this will raise even more puzzles.

\sect{Calabi-Yau manifold with $h_{21} = 2$ as a spectral
hypersurface}

\subsection{Formulation of the problem}

In this section we briefly consider the simplest application
of the prepotential theory - in the version of
section \ref{genpre} - to families of Calabi-Yau manifolds $M$ ($d=3$).
The role of $\Omega$ is played by the holomprphic $(3,0)$-form,
whose existence is peculiar to Calabi-Yau geometry.
Requirement (i) of s.\ref{genpre} is fulfilled if the moduli are
introduced in such a way, that derivatives of
$\Omega$ over them produce singularities on submanifolds of
codimension greater than one.
Requirement (ii)
is fulfilled, since the number of moduli, $h_{30} (M) + h_{21}(M)$
is the same as $\frac{1}{2}{\rm dim} H^3(M)$.
The modulus, associated with $h_{30}(M) = 1$ corresponds
to the overall rescaling of $\Omega$ and is not very
interesting.

The simplest example of Calabi-Yau family,
which is  associated with the $SU(2)$ Seiberg-Witten
theory \cite{KKLMV,KV} has two non-trivial moduli:
$h_{21} = 2$.\footnote{
$h_{21} = {\rm rank}_G +1$, in our case $G = SL(2)$. The unit
difference between $h_{21}$ and $\ {\rm rank}_G$ corresponds
to the dilaton field in the target-space language
and to the $\tau$-parameter (modulus of the {\it bare} spectral
curve) in Whitham terminology.
}
It results from factorization of the $WP^{12}_{1,1,2,2,6}$
submanifold, described by the equation:
\be
p(z|\phi,\psi) \equiv
\frac{z_1^{12}}{12} + \frac{z_2^{12}}{12} +
\frac{z_3^6}{6} + \frac{z_4^6}{6} + \frac{z_5^2}{2}
+ \frac{\phi z_1^6z_2^6}{6} + \psi z_1z_2z_3z_4z_5 = 0.
\ee
The two moduli are $\phi$ and $\psi$.
Relation to $N=2$ SUSY $SU(2)$ pure gauge theory
is originally motivated through the arguments like stringy $S$-duality,
with unbroken gauge group $SU(2)$ emerging at the ''conifold
locus'' in the moduli space,
\be
\left(1 - \frac{\phi}{i\psi^6}\right)^2 + \frac{1}{\psi^{12}} = 0, \ \ \ {\rm
or}
\ \ \ (\phi - i\psi^6 )^2 = 1.
\ee
The periods of $\Omega$ are given by the standard formula:\footnote{
The flat measure restricted to the hypersurface $p = 0\ $ is
$\ dz_1\wedge \ldots\wedge dz_5 \delta(p(z)) = dz_1\wedge \ldots
\wedge dz_5\int d\lambda e^{i\lambda p(z)}$,
and the (quasi)homogeneity of
of $p(z)$ allows one to eliminate $i\lambda$ in the exponent.
}
\be
\Pi_C(\phi,\psi) \equiv
\oint_C dz_1dz_2dz_3dz_4dz_5 e^{p(z|\phi,\psi)}.
\label{CYPeriods}
\ee

Our task could be to demonstrate that the periods $\Pi_C(\phi,\psi)$
satisfy the same Picard-Fuchs equation
(\ref{PicCal}) that the $SL(2)$ Calogero model -
what would allow to conclude that the prepotential
of the {\it stringy} Calabi-Yau model and   that of the simple $1d$
Calogero system are essentially the same.
Instead we obtain a very similar, but essentially different
equation - as one gets beyond the conifold (double scaling) limit.
Elimination or interpretation of this difference remains for
future investigations.

\subsection{Vicinity of the conifold locus: the sine-Gordon
limit of Calogero model}

The double scaling limit in terms of  $\phi$ and $\psi$ is:
$\phi, \psi \rightarrow \infty$ so that  $\phi - i\psi^6$
remains finite.
The correspondence between Calabi-Yau model in this
limit and original Seiberg-Witten example was already established
in \cite{KKLMV}. Near the conifold singularity, the most
 interesting of the periods
(\ref{CYPeriods}) can be reduced (by rescaling of variables and
making a saddle-point calculation)
to the two-fold integrals
\be
\tilde\Pi_C\left(\frac{u}{\Lambda^2}\right) =
\int \frac{dz_1dz_2}{(z_1z_2)^4}
\exp \left(\frac{z_1^{12}}{12} + \frac{z_2^{12}}{12} +
\frac{u z_1^6z_2^6}{6\Lambda^2}\right).
\label{coniperiods}
\ee
Only peculiar combination of moduli,
\be
\tilde u \equiv \frac{u}{\Lambda^2} = \phi - i\psi^6,
\ee
which defines deviation from the conifold locus, survives in this limit.
(The locus itself is described as $u^2 = \Lambda^4$.)
We refer to \cite{KKLMV} for details of  the saddle-point calculation,
leading from (\ref{CYPeriods}) to  (\ref{coniperiods}).
For completeness of our presentation and as a warm-up
for the similar computation beyond conifold limit,
we derive here the Picard-Fuchs equations for $\tilde\Pi(\tilde u)$
and show that they indeed coincide with those for the sine-Gordon
model - i.e. with eq.(\ref{Pic2}).

Denote integration in (\ref{coniperiods})
with the exponential weight  $e^{p(z)}$
by $<\ldots>$, in particular,\footnote{
As usual, Picard-Fuchs equations are for cohomology classes
and do not depend on integration hypersurfaces. For the same
reason we do not need to write $i=\sqrt{-1}$-factors
in the exponents in (\ref{CYPeriods}) and
(\ref{coniperiods}) - they affect only the integration paths and
parametrization of moduli. Of course only for appropriate
integration domains the Ward identites which we use can be
valid. Note also that our notation is slightly different from
\cite{KKLMV}: our $\phi = -\phi_{\rm KKLMV}$ and our
$\psi = -(864i)^{1/6}\psi_{\rm KKLMV}$ so that parameters
$X = \frac{\phi}{i\psi^6}$, $Y = \frac{1}{\phi^2}$,
$X_1 = \frac{1}{(\phi - i\psi^6)^2}$,
$X_2 = \frac{\phi - i\psi^6}{\psi^6} =
-i\left(1 - \frac{\phi}{i\psi^6}\right)$.
}
$\tilde\Pi(\tilde u) = \ < (z_1z_2)^{-4}>$.
There always is a Ward-identity, resulting from the change of
integration variables $z_j$:
\be
<\ F(z)\frac{\partial p(z)}{\partial z_j}\delta z_j
+  F(z)\frac{\partial \delta z_j}{\partial z_j} +
\frac{\partial F}{\partial z_j}\delta z_j\ >\ = 0.
\label{WIgen}
\ee
for any function $F(z)$.
Whenever $\delta z_j$ is adjusted in such a way that the operator
at the l.h.s. can be obtained by action of some differential operator
on $< (z_1z_2)^{-4} >$ - we get a Picard-Fuchs equation.
In the case of (\ref{coniperiods}) the simplest option is
\be
\delta z_1 = z_1z_2^{12}, \nn \\
\delta z_2 = \left( 3 - \tilde u(z_1z_2)^6\right)z_2
\ee
($\delta z_2$ is adjusted to cancel all the terms associated with
$\delta z_1$, which are not integer powers of $z_1^6z_2^6$).
With this choice and with $F(z) = (z_1z_2)^{-4}$ (\ref{WIgen}) turns into
\be
0 = \ < (1-\tilde u^2)z_1^{12} z_2^{12}  - 9 >\ =
\left(36(\Lambda^4 - u^2)\frac{\partial^2}{\partial u^2} - 9
\right)\tilde \Pi\left(\frac{u}{\Lambda^2}\right)
\label{Pic2froCY}
\ee
what is  exactly the Picard-Fuchs equation (\ref{Pic2}) for
the sine-Gordon model.

\subsection{Beyond conifold limit}

We can now apply the same trick to derive the Picard-Fuchs
equation for $\Pi_C(\phi,\psi)$ beyond conifold limit.
Now we use $<\ldots>$ to denote integration in
(\ref{CYPeriods}), so that $\Pi(\phi,\psi) = \ <1>$.
This time we substitute into (\ref{WIgen}) $F(z) = 1$ and
\be
\delta z_1 = z_1z_2^{12}, \nn \\
\delta z_2 = -\left( 1 + \phi(z_1z_2)^6 + \psi z_1z_2z_3z_4z_5\right)z_2
\ee
 and obtain:
\be
0 = \ < \ (1-\phi^2)z_1^{12}z_2^{12} - 8\phi z_1^6z_2^6 - 1  \nn \\
 - 2\phi z_1^6z_2^6 \cdot  \psi z_1z_2z_3z_4z_5 -
3\psi z_1z_2z_3z_4z_5 - (\psi z_1z_2z_3z_4z_5)^2\ >
\ee
Obviously, the r.h.s. can be reproduced by moduli-derivatives.
After some arithmetics we obtain the Picard-Fuchs equation in the form:
\be
\left({\cal D}_1^2 - ({\cal D}_2+\frac{1}{6})^2\right)\Pi = 0,
\ \ \ {\rm or}\ \ \
\left({\cal D}_1^2 - {\cal D}_2^2\right)(\psi\Pi) = 0
\label{ABPi}
\ee
where operators ${\cal D}_{12}$ are given by\footnote{
Remarkably, the algebra formed by these operators,
$\left[{\cal D}_1, {\cal D}_2\right] = {\cal D}_1 $, has another well
known representation: in terms of $p$ and $q$, satisfying $[p,q]=1$:
$\hat{\cal D}_1 = e^q$, $\hat{\cal D}_2 = p$ - so that
(\ref{ABPi}) becomes the Shroedinger equation for $1d$ Liouville
model: $(p^2 - e^{2q})\Pi = 0$. This important analogy is, however,
beyond the scope of this paper. For the relevant theory of
Liouville wave functions see \cite{Liou} and references therein.
}
\be
{\cal D}_1 = \frac{\partial}{\partial \phi}, \ \ \ \
{\cal D}_2 = \phi\frac{\partial}{\partial \phi} +
\frac{1}{6}\psi\frac{\partial}{\partial \psi}.
\ee
Note, that this equation is essentially insensitive to the part
of defining polynomial $p(z|\phi,\psi)$ which depends on $z_3$, $z_4$
and $z_5$: it only matters that they all interact with $z_1$ and $z_2$
through the linear term $\psi z_1z_2 f(z_3,z_4,z_5)$ and the coefficient
$\psi$ is a modulus.
Of course, as in the $SL(2)$ Calogero case,
since we are now in the situation with two moduli, the
second Picard-Fuchs equation is also needed.  But it should be higher
order in derivatives and does not add to our discussion.

\subsection{The double scaling limit of Picard-Fuchs equation}

As we already know, in this limit $\phi$ and $\psi \rightarrow \infty$,
while $\tilde u \equiv \frac{u}{\Lambda^2} = \phi - i\psi^6$ remains
finite. Moreover,
\be
\tilde\Pi = \psi^{-4}\Pi,
\ee
this can be shown
either by saddle-point evaluation of $z_{3,4,5}$-integrals in
(\ref{CYPeriods}) - what gives (\ref{coniperiods}) with an extra
factor of $\psi^{-4}$, or by comparison of the equation that we obtain
shortly with (\ref{Pic2}). For such $\tilde\Pi$ the Picard-Fuchs
equation (\ref{ABPi}) becomes:
\be
\left({\cal D}_1^2 - \left({\cal D}_2 - \frac{1}{2}\right)^2\right)
\tilde\Pi = 0, \ \ \ {\rm or} \ \ \
({\cal D}_1 - {\cal D}_2)({\cal D}_1 + {\cal D}_2 - 1)\tilde\Pi =
\frac{1}{4}\tilde\Pi.
\label{ABPii}
\ee
Perform now a change of variables: $\{\phi,\psi\} \rightarrow
\{\phi,u\}$. Then,
$$
\left.\frac{\partial}{\partial \phi}\right|_{\psi = {\rm const}} =
\left.\frac{\partial}{\partial \phi}\right|_{u = {\rm const}} +
\Lambda^2\frac{\partial}{\partial u}, \ \ \ \
\frac{\partial}{\partial\psi} = -6i\Lambda^2\frac{\partial}{\partial u}.
$$
After such substitution (\ref{ABPii}) turns into:
\be
\left(\frac{\partial}{\partial\phi} + \Lambda^2\frac{\partial}{\partial u}
\right)^2\tilde\Pi =
\left(\phi\frac{\partial}{\partial\phi} + u\frac{\partial}{\partial u}
- \frac{1}{2}\right)^2\tilde\Pi,
\ee
and assuming $\frac{\partial\tilde\Pi}{\partial\phi}=0$ for the leading
term in the $\frac{1}{\phi}$ expansion of $\tilde\Pi$, we obtain
(\ref{Pic2}), or, what is the same, (\ref{Pic2froCY}):
\be
\left( (u^2 - \Lambda^2)\frac{\partial^2}{\partial u^2} + \frac{1}{4}\right)
\tilde\Pi = 0.
\ee

\subsection{Comments \label{DisCY}}

Equation (\ref{ABPi}) is similar enough to eq.(\ref{PicCal}) for $1d$
Calogero model to confirm the belief that the two are closely
related, though  the  equations are clearly
not the same (this is especially obvious if Calogero-case
equation is convertable into something like (\ref{PicCaldream})).
The most striking difference is appearence of two second-order
differential operators in (\ref{ABPi}), while one of them is of the
first order in (\ref{PicCal}). It deserves emphasizing that this
first order derivative is exactly what makes (\ref{PicCaldream}) look
like a Schroedinger equation - and thus it is not very easy to give up.
This discreapancy - if not eliminated by a clever change of variables,-
deserves explanation. If eliminated (and if the second Picard-Fuchs
equations are also compatible), it would strongly suggest that
the stringy Calabi-Yau and $1d$ Calogero belong to the same
Whitham universality class - not a big surprise, because the
Whitham theory classifies models according to the number of their
moduli: extreme low-energy degrees of freedom rather than to
that of ordinary degrees of freedom at  high energies and models
which are very different in the UV can become the same in the IR.
Most probably, however, the difference between (\ref{ABPi}) and
(\ref{PicCal}) is real, and disappears only in the limit $\alpha '
\rightarrow 0$. It is very appealing to think that (\ref{ABPi}) is
a ''relativistic'' version of (\ref{PicCal}) - when the linear
''time''-derivative turns into  the second-order one. If correct, this
would strongly imply that the Ruijsenaars model - the ''relativistic''
version of Calogero system - can adequately describe (\ref{ABPi}).
For our purpose this model is just the same as Calogero one, just the
group $G$ ($G = SL(2)$ in most of our examples) is deformed to become
quantum group, $G_s = SL_s(2)$, with the ''Planck-constant''
$\ \log s \sim \alpha'$. Unfortunately, our prelimiary analysis in
s.\ref{Rumo} below does not fully confirm these expectations. It can be,
however, too naive in its present form.

Another remark: there are $6$ periods on the Calabi-Yau manifold
in question, and only two of them can be related to $a$ and $a^D$
of Calogero model. The others are similarly related to the
''trivial'' periods $(1,\tau)$ of $E(\tau)$, which are neglected when
${\cal C}$ is substituted by $\hat{\cal C}$, see eq.(\ref{hatC}).
This statement is in agreement with the results of \cite{KKLMV}:
the other $4$ periods in the double scaling limit are $(1,S,u,uS)$,
and in this limit $S \sim \tau$.

\sect{Ruijsenaars model: beginning of the story \label{Rumo}}

Since this model have not been analyzed in the context
of Seiberg-Witten theory in \cite{IM}, we need several extra
formulas to begin with. The Lax matrix for Ruijsennars model
- which we need to define the full spectral curve ${\cal C}$
from eq.(\ref{curve}) - is \cite{Ru}:
\be
L_{ij}(\xi) = e^{P_i} \frac{F(q_{ij}|\xi)}{F(q_{ij}|\mu)}
\prod_{l\neq i} n(\mu)\sqrt{\wp(\mu) - \wp(q_{il})},
\label{RuLax}
\ee
where $i,j,l = 1,\ldots, N_c$, $\ \wp(\xi) = -\frac{\partial\zeta(\xi)}
{\partial \xi} = - \frac{\partial^2\log\sigma(\xi)}{\partial \xi^2}$,
$$
F(q_{ij}|\xi) = \frac{\sigma(q_i-q_j + \xi)}{\sigma(\xi)\sigma(q_i-q_j)},
$$
and normalization constant $n(\mu)$ can be fixed as convenient,
most often it is taken to be $n(\mu) = \sigma(\mu)$ - this simplifies
the formulas in the Calogero limit $\mu \rightarrow 0$.
The {\it bare} spectral curve - to which the spectral parameter
$\xi$ belongs in eq.(\ref{RuLax}) is our usual elliptic $E(\tau)$.

$L_{ij}(\xi)$ in (\ref{RuLax}) depends on an extra parameter $\mu$,
as compared to Calogero model, and Calogero Lax operator
\cite{KriCaLax} is recovered in the limit $\mu \rightarrow 0$,
$P_i = \frac{\mu}{g}p_i + o(\mu^2)$ \cite{Ru}:
\be
L_{ij}(\xi) = \delta_{ij} \ +\ \frac{\mu}{g}\left(p_i\delta_{ij} +
(1-\delta_{ij})gF(q_{ij}|\xi)\right) \ +\  o(\mu^2).
\ee

The spectral curve ${\cal C}$ for the Ruijsenaars model, as defined
in (\ref{curve}), is:
\be
{\cal C}: \ \ \ \ \ \det \left(t \delta_{ij} - L_{ij}(\xi)\right) = 0,
\label{Rucurve}
\ee
and according to (\ref{diff}),
\be
d{\cal S}_{\rm min}^{Ru} = 2td\omega_0(\xi).
\ee

We now restrict ourselves to our usual example of $G = SL(2)$, i.e.
$N_c = 2$. Then (\ref{Rucurve}) is
\be
t^2 - t{\rm tr L} + \det L = 0
\ee
and, denoting $ P \equiv P_1 = -P_2$, $q \equiv q_1 - q_2$,
\be
H \equiv \frac{1}{2n(\mu)}{\rm tr} L = \frac{1}{2} (e^P + e^{-P})
\sqrt{\wp(\mu) - \wp(q)}, \nn \\
\det L = n^2(\mu)\left(\wp(\mu) - \wp(\xi)\right),
\ee
(to get these formulas one can use the identity
$F(q|\xi)F(-q|\xi) = \wp(q) - \wp(\xi)$, see \cite{IM} for generalization
to higher-$N_c$ case).
$H = H(\mu)$ that appeared here is the single independent integral
of motion  of the $SL(2)$-Ruijsenaars model. In terms of it, we rewrite
(\ref{Rucurve}) as:
\be
t = \frac{H \pm \sqrt{H^2 - \wp(\mu) + \wp(\xi)}}{n(\mu)},
\ee
and
\be
n(\mu) d{\cal S}_{\rm min}^{Ru} = 2Hd\omega_0 \pm
\left.d{\cal S}_{\rm min}^{Cal} \right|_{\hat h = H^2 - \wp(\mu)},
\ee
where $d{\cal S}_{\rm min}^{Cal} = 2\sqrt{\hat h + \wp(\xi)}d\omega_0
\sim
\frac{\sqrt{\hat h - x}}{y(x)}dx$, with $\hat h = H^2(\mu) - \wp(\mu)$.

We see immediately that the substitution (projection)
${\cal C} \rightarrow \hat{\cal C}$ (\ref{hatC}), that played a certain
role in our analysis of Calogero model, is a little more tricky in
the context of Ruijsenaars model. Actually, it is only
$d{\cal S} \equiv n(\mu)d{\cal S}_{\rm min}^{Ru} - 2Hd\omega_0$
that can be projected on $\hat{\cal C}$ without loss of information
about the periods. Instead, the piece $2Hd\omega_0$ keeps track
of the periods of $E(\tau)$ - which are completely ignored by
$d{\cal S}_{\rm min}^{Cal}$. This is in accordance with one of our
expectations in s.\ref{DisCY} - that more periods can be revealed
in the Ruijsenaars model context - as required to make things
consistent  with Calabi-Yau picture.

However, the Picard-Fuchs equation does not change as much as
necessary. Essentially, $d{\cal S}$ satisfies just the same
equation (\ref{PicCal}): the only new thing is that one should
subtract $2Hd\omega_0$ and change the variables. Eq.(\ref{PicCal})
is written in variables $(\hat h,\tau)$, while now we rather need
$(H,\tau)$. Since
$$
\left.\frac{\partial}{\partial \tau}\right|_H =
\left.\frac{\partial}{\partial\tau}\right|_{\hat h} +
\frac{\partial \hat h}{\partial H}\frac{\partial}{\partial\hat h}, \ \ \ \
\frac{\partial}{\partial H} = \frac{\partial\hat h}{\partial H}
\frac{\partial}{\partial\hat h}
$$
this change of variables does not seem to produce second-order
$\tau$-derivatives and thus can hardly eliminate the main
discrepancy with the Calabi-Yau-case equation (\ref{ABPi}).

Still, this analysis of the Ruijsenaars model is preliminary
and can appear oversimplified. In any case the problem deserves
attention and requires more studies in the future.

\sect{Open questions}

Before concluding this paper we briefly list  the problems
that did not allow us to make the presentation completely closed.

As concerns the abstract theory of the prepotential, discussed in
the first half of the paper,

- Our presentation was essentially based on the use of the
''quasiclassical Baker-Akhiezer function'' $dS$ (or $\Omega$
in sec.\ref{genpre}). Quasiclassical $\tau$-function (of which
the prepotential is a logarithm) appears as a secondary object.
The proper presentation should be inverse: the quasiclassical
$\tau$-function should be introduced first, and Baker-Akhiezer
function emerge from it. In the theory of ordinary $\tau$-functions
these are defined as generating functionals of all the matrix
elements of a universal group element of some Lie algebra, and
- as such - satisfy the (generalized) bilinear Hirota equations
(which are nothing  but comultiplication formula, see \cite{GKLMM}
and references therein).  In the particular case of level-{\it one}
affine (Kac-Moody) algebras - when conventional
(multicomponent) KP/Toda hierarchies arise - one can proceed
further: define the
Baker-Akhiezer function by eq.(\ref{BAf}) and it will be a function
on (section over) a complex {\it curve}. Even for the ordinary
$\tau$-functions the substitute of Miwa transform and thus of
eq.(\ref{BAf}) in the general situation (for arbitrary Lie groups,
not obligatory level-one and 1-loop) is unknown.
It is suggested in \cite{D} that analogous abstract definition
of quasiclassical $\tau$-functions should be based on the concept
of Frobenius, rather than Lie algebras - but these one are much
less understood, to say the least. In any case, emergence of
$d{\cal S}$ - and algebraic criteria of when it arises as living on
the spectral {\it curves} and when on some higher-dimensional
manifolds (i.e. when one should use $\Omega$ from sec.\ref{genpre}
above instead of
$d{\cal S}$) - in such context remain unclear.

- Relation between group theory (ordinary $\tau$-functions)
and Hodge structures (quasiclassical ones) -
which in physical  language should be  relation between the
entire model and its topological (low-energy) limit - is still
obscure. Of course it is well understood in the particular context
of Whitham limit of KP/Toda $\tau$-function (see \cite{NT}
and references therein), but again, in general abstract terms
things are not enough clarified.

- Conceptual relation between the prepotential, considered as a full
generating function (i.e. with infinitely many "time"-variavles),
and the reduced one, defined with the help of particular $d{\cal S}$
or $\Omega$ should be understood better: this is just another
formulation of the previous problems.

- Even in the setting to which we had to restrict our presentation,
there remains a small subtlety: divergence of the prepotential in
the presence of simple poles in $d{\cal S}$ and $\Omega$
(in the Seiberg-Witten theory these arise whenever there are
massive matter multiplets).  Though not very harmfull - as explained
in the main text - regularization of these divergencies is an
additional step in the construction (which can also introduce
extra moduli, characterizing the choice of regularization),
which should be made somewhat less artificial than it is now.

As concerns the second part of the paper: examples
of how the general theory can be applied to particular
Seiberg-\-Witten/integrable models, we mostly concentrated
on the $SL(2)$ case. Thus,

- Generalization to other groups, most interestingly to {\it quantum} and
{\it affine} groups, is not discussed. It is rather straightforward,
though sometime technically tedious, especially in the case
of Calabi-Yau manifolds. Adequate technical means should be
developed to handle generic situation, especially the
Picard-Fuchs equations and Calogero
($N=2 \rightarrow N=4$ flow) case.

- A separate issue is adequate treatement of models with matter
multiplets in the fundamental of the gauge group. This is not
a really big problem at the first step, because it is enough
to allow more punctures on the bare spectral curve in order
to get {\it some} description, but the next steps - like embedding
of the model into an UV-finite one (say, the substitute
of Calogero-model description for the $N_f=4$ model) - can
hardly be made without adequate techniques.

Finally, even for the $SL(2)$ case we did not fully
resolve three kinds of problems:

- We did not treat the modulus $\tau$ in Calogero system
on equal footing with the other arguments of the prepotential:
there is no ''symmetry'' between $\hat h$ and $\tau$.

- A full understanding of Picard-Fuchs equation (\ref{PicCal})
in Calogero case is not achieved: see ss.\ref{CafPF} and
\ref{CoPic} for details.

- Relation between Calabi-Yau and Calogero models is not
completely revealed, see discussion in s.\ref{DisCY}.
Once found it could be an important step towards classification
of prepotentials in the simplest non-trivial case of two moduli.

To this list of ''technical'' problems one should of course add
a string of conceptual ones from \cite{Go}: At one end would be
understanding of Whitham theory as a general method for
description universality classes of renormalization-group flows
and - at the same time - of quantization of the field theory.
At another end  would be classification of all possible Witham
universality classes - possible if the theory is formulated in
abstract terms. When worked out and then combined
the answers to these conceptual questions will definitely have
interesting  applications.

\sect{Conclusion}

To conclude, we presented a preliminary discussion of quasiclassical
$\tau$-functions, or prepotentials - the old-known objects, which
nowadays acquire the role they deserve in theoretical
physics. As ordinary (generalized) $\tau$-functions are supposed
to describe generic effective actions in the space of coupling
constants and exhibit their intrinsic group-theoretical nature,
the quasiclassical ones should play the same role at the end-point
of renormalization-group evolution: in the IR limit - and the
corresponding theory should reveal the hidden dynamical
symmetry of effective models on the moduli spaces. Unfortunately,
this theory at the present stage is not so clear as in the case of
the usual $\tau$-functions: it more resembles the ''phenomenology''
of  $\tau$-functions before they were identified with universal
group elements.
By now standard identification of the prepotential (Whitham)
theory with that of the Hodge structures - which is used
in the present paper  -  is at best the first step towards the
general theory. As a necessary minimum, one should clearly
explain, why (and in what sense) the Hodge structures
{\it always} appear at the IR stable points of renormalization
evolution of {\it group theory} (by now it is a problem even to
define what these {\it words} could mean in the absolutely general
setting).

Still, even with the present state of knowledge, the theory has
applications - of which we briefly discussed the most fresh
and exciting ones: to Seiberg-Witten description of $N=2$
SUSY gauge models and its lifting to Calabi-Yau models
(the yet-remembered previous avatara was in topological
field theory). We paid some attention to the homogeneity
property, peculiar to {\it all} prepotentials - which reflects
its invariance under
the simplest constituent of renormalization-group evolution
- (in)dependence on the overall scale. As soon as some
moduli are fixed, the reduced prepotential acquires anomalous
dimension, defined in the Seiberg-Witten setting by its
inclusion into anomaly multiplet,\footnote{
$\beta_W$ is Wilsonian $\beta$-function, which is always
constant (one-loop) in supersymmetric gauge theories
in the absence of massive matter fields. In one word, the
field-theory derivation of (\ref{ano}) is just the same as the
usual one for the trace anomaly: Consider effective action
$$
\exp \int{\cal F}(\Phi_{\rm cl})d^4xd^4\theta =
\int {\cal D}\Phi_{\rm qu} \exp \int F(\Phi_{\rm cl} +
\Phi_{\rm qu}) d^4xd^4\theta \equiv \ <1>_{\Phi_{\rm cl}}
$$
and vary with respect to the scale $\Lambda$. Then,
since the prepotential at the l.h.s. depends on $\Lambda$
only through $\phi_{\rm cl}/\Lambda$,
$$
\int\left(2{\cal F} - \sum_i \Phi_i\frac{\partial{\cal F}}{\partial
\Phi_i}\right)\delta\Lambda\ d^4xd^4\theta =
\ \langle\langle \int \frac{\partial F}{\partial\Lambda}
\delta\Lambda d^4xd^4\theta\rangle\rangle
$$
If one wants to recover the axial-current anomaly from the
same derivation, one formally allows for complex variations
$\delta\Lambda$. Similarly, allowing  nilpotent $\delta\Lambda$
we  obtain this relation for all the components of the superfield,
not just for the highest one. Since the highest component of
$\frac{\partial F}{\partial \Lambda}$ is $\beta_W{\rm tr}\ G^2$,
the lowest one is obviously $\beta_W{\rm tr}\ \Phi^2$.
Comparing  this with the lowest component at the l.h.s. we get
(\ref{ano}). More accurately, one should couple the theory
to the $N=2$ supergravity and vary w.r.to rescaling of the
gravitational supermultiplet. Also one should treat auxiliary
fields more carefully (note that the bare prepotential $F$
is non-abelian and therefore can be described only with
inclusion of infinitely many auxiliary superfields).
}
along with
$T_{\mu\mu} \sim \beta_W<{\rm tr}\ G^2>$ and
$\partial_\mu J_\mu^5 \sim \beta_W <{\rm tr}\  G\tilde G>$:
\be
2{\cal F} - \sum_i a_i\frac{\partial{\cal F}}{\partial a_i} \sim
\beta_W<{\rm tr}\ \phi^2>
\label{ano}
\ee
(all the relations acquire corrections when massive matter fields
are introduced). This simple example illustrates the very spirit
of the general theory (or rather entire ''stringy'' philosophy):
as soon as all  possible parameters (including IR and UV
cut-offs, counterterms, bare couplings etc) are introduced
in the game as either coupling constants or background fields, -
the hidden symmetries are revealed, and they often do have
implications for original reduced model.

Clearly, much more should be done, in order to appreciate
the real significance of the string program for physics, as to  the
examples considered in these notes. They can hopefully add
to the evidence that the Whitham theory is already a
full-fledged constituent of the whole approach.

\section{Acknowledgements}

We benefited from discussions with
O.Aharony,  L.Alvarez-Gaume, S.Das, A.Gera\-simov, C.Gomez,
A.Gorsky, A.Hanany, A.Losev, I.Kri\-che\-ver,
H.Ku\-ni\-to\-mo,
A.Mar\-shakov, A.Mironov,  N.Nekrasov, K.Oh\-ta, M.Ol\-sha\-netsky,
I.Polyubin, A.Rosly,  V.Rubtsov, J.Sonnenschein and A.Tokura.

H.I. is supported in part by Grant-in-Aid for Scientific Research
(07640403) from the Ministry of Education, Science and Culture, Japan.
A.M. acknowledges the
hospitality of  Osaka University and support of the JSPS.

\end{document}